\documentclass{elsart}

\input epsf

\begin{document}

\begin{frontmatter}

\title{Apparatus for a Search for T-violating Muon Polarization 
                    in Stopped-Kaon Decays}

\author[tsukuba,kek]{M.~Abe,\thanksref{second}}
\author[kek,osaka]{M.~Aoki,\thanksref{second}}
\author[tsukuba]{I.~Arai,}
\author[tsukuba]{Y.~Asano,}
\author[sask]{T.~Baker,}
\author[vpi]{M.~Blecher,}
\author[kek]{M.D.~Chapman,}
\author[montreal]{P.~Depommier,}
\author[triumf]{P.~Gumplinger,}
\author[ubc]{M.~Hasinoff,}
\author[triumf]{R.~Henderson,}
\author[tsukuba,kek]{K.~Horie,\thanksref{second}}
\author[tsukuba,kek]{Y.~Igarashi,\thanksref{second}}
\author[tsukuba,ipcr]{T.~Ikeda,\thanksref{second}}
\author[kek]{J.~Imazato,}
\author[inr]{A.P.~Ivashkin,}
\author[kanagawa]{A.~Kaga}
\author[yonsei]{J.H.~Kang,}
\author[inr]{M.M.~Khabibullin,}
\author[inr]{A.N.~Khotjantsev,}
\author[korea]{E.J.~Kim}
\author[korea]{K.U.~Kim}
\author[inr]{Y.G.~Kudenko,}
\author[kek,osaka]{Y.~Kuno,\thanksref{second}}
\author[yonsei,kriss]{J-M.~Lee,\thanksref{second}}
\author[korea]{K.S.~Lee,}
\author[korea,kek]{G.Y.~Lim,\thanksref{second}}
\author[triumf]{J.A.~Macdonald,\thanksref{coraut}}
\author[princeton]{D.R.~Marlow,}
\author[princeton]{C.R.~Mindas,}
\author[inr]{O.V.~Mineev,}
\author[sask]{C.~Rangacharyulu,}
\author[tsukuba]{S.~Sekikawa,}
\author[tsukuba]{K.~Shibata,}
\author[kek,fnal]{H.M.~Shimizu,\thanksref{second}}
\author[osaka]{S.~Shimizu,}
\author[sask]{Y.-M.~Shin,}
\author[yonsei]{Y.H.~Shin,}
\author[korea]{K.S.~Sim,}
\author[tsukuba]{A.~Suzuki,}
\author[tsukuba]{T.~Tashiro,}
\author[tsukuba]{A.~Watanabe,}
\author[triumf]{D.~Wright,}
\author[korea]{C.H.~Yim}
\author[kek]{T.~Yokoi}
~\\
(E246 KEK-PS COLLABORATION)

\address[tsukuba]{University of Tsukuba, Tsukuba, 305-0006 Japan}
\address[kek]{High Energy Accelerator Research Laboratory (KEK), Tsukuba, 305-0801 Japan}
\address[inr]{Institute for Nuclear Research of RAS, 117312 Moscow, Russia}
\address[sask]{University of Saskatchewan, Saskatoon, S7N 5E2 Canada}
\address[vpi]{Virginia Polytechnic Institute and State University, 
Blacksburg VA 24061-0435, U.S.A.}
\address[montreal]{Universit\'{e} de Montr\'{e}al, Montr\'{e}al, H3C 3J7 Canada}
\address[triumf]{TRIUMF, Vancouver, V6T 2A3 Canada}
\address[ubc]{University of British Columbia, Vancouver, V6T 1Z1 Canada}
\address[ipcr]{The Institute of Physical and Chemical Research, Saitama, 351-0198 Japan}
\address[yonsei]{Yonsei University, Seoul, 120-749 Korea}
\address[kanagawa]{Kanagawa University, Hiratsuka, 259-1293 Japan}
\address[korea]{Korea University, Seoul, 136-701 Korea}
\address[princeton]{Princeton University, Princeton NJ 08544, U.S.A.}
\address[osaka]{Osaka University, Osaka, 560-0043 Japan}
\address[fnal]{Fermilab Computing Division, Batavia IL 60510-0500, U.S.A.}
\address[kriss]{Korea Research Institute of Standards and Science, Taejon, 305-346 Korea}

\thanks[second]{Second address is the present address.}
\thanks[coraut]{Corresponding author. email: jam@triumf.ca}

\begin{abstract}

The detector built at KEK to search for T-violating transverse muon polarization in $K^+ 
\rightarrow \pi^0 \mu^+ \nu_{\mu}$ ($K_{\mu3}$) decay of stopped kaons 
is described.
Sensitivity to the transverse polarization component is obtained from reconstruction of the decay plane by tracking the $\mu^+$ through a toroidal spectrometer and detecting the $\pi^0$ in a segmented CsI(Tl) photon calorimeter. The muon polarization was obtained from the decay positron asymmetry of  muons stopped in a polarimeter. The detector included features which minimized systematic errors while maintaining high acceptance.

\end{abstract}


\end{frontmatter}

\section{Introduction}
\label{sec:intro}

Violation of time-reversal invariance (T) could be 
inferred from a weak interaction induced transverse 
polarization ($P_T$) of muons normal to
the decay plane in the decays
$K^+\rightarrow\pi^0\mu^+\nu$~($K_{\mu3}$) and  
$K^+\rightarrow\mu^+\nu\gamma$~($K_{\mu2\gamma}$).
Standard Model (SM)
contributions to $P_T$ are at about the $10^{-7}$ level \cite{valencia}
 and final state interactions in $K_{\mu3}$ contribute at the $<10^{-5}$ level \cite{fsi}.
A measurement in excess of this has
the potential to  reveal new CP violation physics, given CPT invariance.
The value of $P_T$ can be as large a $10^{-3}$ -- $10^{-2}$ in models 
with multi-Higgs doublets, leptoquarks, left-right symmetry or SUSY \cite{nonsm}.
Motivation to search for additional sources of CP violation arises from the
observed baryon asymmetry in the universe, which cannot be explained by the CP
violation in the Standard Model alone such as the SM T-violation
 observed in the neutral kaon system  \cite{cplear}. 
At the High Energy Accelerator Research
Organization (KEK) in Japan, the E246 collaboration
 is completing a search for $P_T$  in $K_{\mu3}$ at the $\sim 10^{-3}$
level.

In 1999, the first result, $P_T= -0.0042 \pm 0.0049 (stat) \pm 0.0009 
(syst)$, was published \cite{fstrslt}, based on 
$\sim 3.9 \times 10^6$ good 
$K^{+}_{\mu3}$ events from the 
data taken during 1996 and 1997, indicating there is no evidence for T violation.
 The T-violating parameter Im$\xi$  was then 
extracted as Im$\xi= -0.013 \pm 0.016 (stat) \pm 0.003 (syst)$.
 The 90\% confidence limits are given as 
$\vert P_T \vert < 0.011$ and $\vert {\rm Im}\xi \vert < 0.033$.
Further data were collected during 1998 -- 2000, and 
the final E246 sensitivity to $P_T$ in $K_{\mu3}$ is expected to be
 $\sim 2\times 10^{-3}$ 
 corresponding 
  to $\delta$Im$(\xi)$ of about
$7\times 10^{-3}$. Since statistics will dominate the sensitivity,
the detector will have
 a capability to  improve the sensitivity 
 by a factor of 3--4 using a more intense kaon beam in the future.
In this paper we give a description of the apparatus.

\section{Experimental Method and Detector Overview}
\label{sec:overview}

\subsection{Principle of measurement}
\label{subsec:princmeas}

The measurement of the T-violating transverse muon polarization 
$P_{T}$ requires 
first that the detectable constituents of the $K_{\mu3}$ decay with 
a branching ration of 3.2\% be distinguished 
and that the 
decay plane be established from the kinematic parameters. Secondly, a 
polarimeter must be sensitive to the polarization component normal to 
the decay plane. For example, if events are selected whose decay plane includes 
the beam axis, then $P_{T}$ would appear as an azimuthal
component of the polarization around the beam axis. The muon 
polarization is measured from the
asymmetry of the decay positron angular distribution
relative to the polarization axis.

\subsection{Stopped beam method}
\label{subsec:stoppedk}

In our experiment we used a stopped kaon beam in contrast to
the previous search  \cite{blatt} which
used the in-flight decay of a 4-GeV kaon beam. Although a low-momentum 
stopped-kaon beam risks a potential disadvantage of reduced 
kaon flux compared to the high momentum beam used for in-flight 
measurements, the stopped method 
offers several potential 
advantages in minimizing backgrounds and systematic uncertainties 
which compensate the reduced beam intensity, and can result in a measurement 
with improved overall sensitivity.

In a stopped-kaon decay, the kinematics can be determined with high 
resolution which allows a precise definition of the decay 
plane as well as powerful tools to reject backgrounds and reduce 
systematic instrumental sources of asymmetry. In our 
apparatus, muons are accepted in a range of polar angles around $90^{\circ}$ 
with respect to the beam axis and are momentum analyzed in one 
of the 12 gaps of a superconducting toroidal spectrometer magnet. 
The decay vertex is determined by tracking in 
scintillators and/or wirechambers of the charged incoming kaon and 
outgoing muon. The muon momentum is obtained 
with adequate resolution to reject the dominant decay modes 
$K^+\rightarrow\pi^+\pi^0$ ($K_{\pi2}$) 
and $K^+\rightarrow \mu^+\nu$ ($K_{\mu2}$).
The $\pi^{0}$ momentum is obtained from a high-solid-angle segmented 
CsI(Tl) calorimeter, and the kaon momentum is, of course, zero. 

\begin{figure}
\epsfxsize=\linewidth
\begin{center}
\epsffile{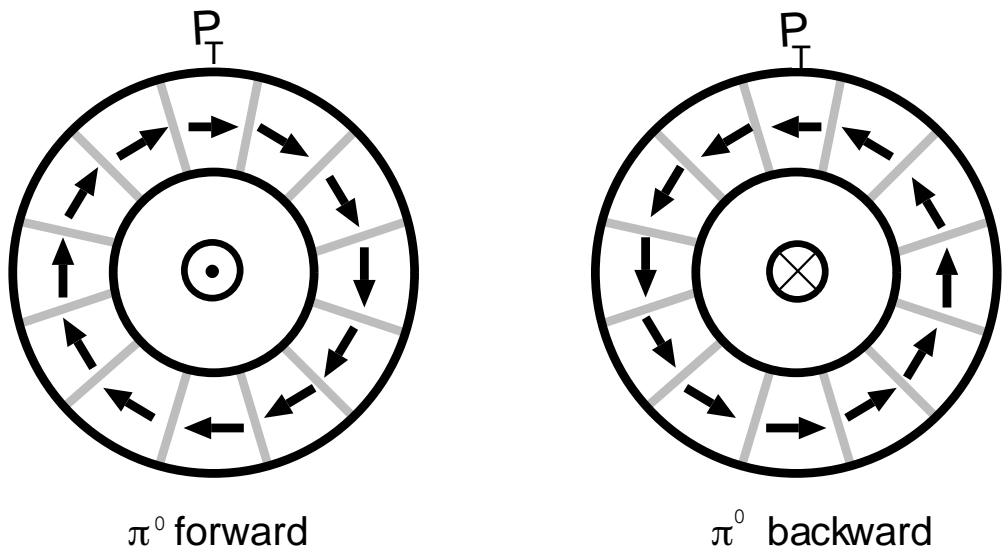}
\end{center}
\caption{Comparison of $P_T$ for $fwd$ and $bwd$ going $\pi^0$s.}
\label{fg:fwdbwdpt}
\end{figure}

Good determination of the $\pi^{0}$ decay kinematics in the segmented 
calorimeter allows one to identify and include 
events in the muon polarimeter corresponding to both forward (downstream) and backward (upstream)
going $\pi^{0}$s, $fwd$ and $bwd$ events, respectively.
The two classes of 
events 
give opposite asymmetries in the polarimeter due to $P_{T}$ as 
illustrated in Figure \ref{fg:fwdbwdpt}. That is, a properly 
normalized sum of $fwd$- and $bwd$-type events must give a zero 
asymmetry.
In the previous in-flight experiment only a
forward-decaying ($fwd$)
$\pi^{0}$ was inferred from the forward boost from the $K$ rest frame 
and an energy threshold on one of the decay photons.

From the kinematics it is also possible to optimize the instrumental 
sensitivity to  the physics parameter of T violation in this decay,
${\rm Im}\xi$. The value of $\langle P_{T} / {\rm 
Im}\xi \rangle$ varies by an order of magnitude over the 
$\mu^{+},\pi^{0}$ phase space. A figure of merit combining the 
statistical sensitivity of the decay rate 
and  $\langle P_{T} / {\rm Im}\xi \rangle$ peaks for $\mu^{+}$ 
energies above 50 MeV and $\pi^{0}$ energies below $\sim 80$ MeV, well 
away from the kinematic peak. The 
muon acceptance in particular can therefore be optimized to enhance 
the sensitivity to ${\rm Im}\xi$ from $P_{T}$.

\subsection{Longitudinal field and double ratio method}
\label{subsec:longfield}

\begin{figure}
\epsfxsize=\linewidth
\begin{center}
\epsffile{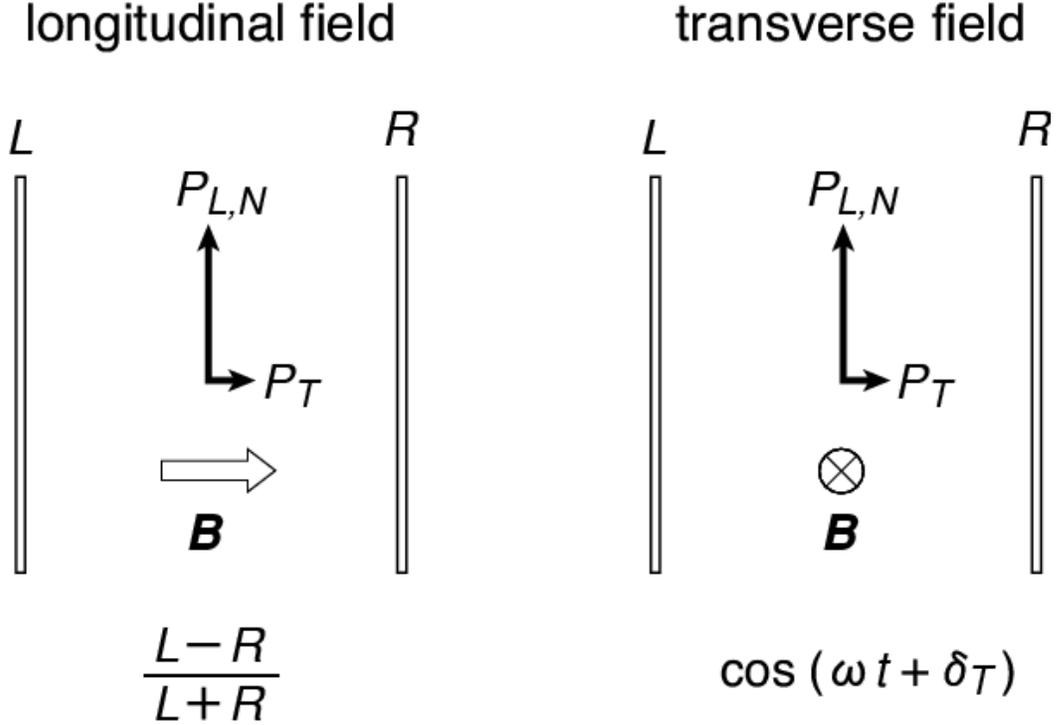}
\end{center}
\caption{Comparison of longitudinal and transverse holding field methods.}
\label{fg:msr}
\end{figure}

Instead of using the transverse
field method which was adopted in the previous
experiment\cite{blatt} we adopted the longitudinal field configuration.
In this case the transverse component $P_T$ of the muon polarization is held, and the non-T-violating in-plane components precess, while in the transverse field method the in-plane 
component $P_L$ parallel to the muon momentum is held and the T-violating
component was sought as a possible phase shift of the precessing
component. The comparison is shown in Figure \ref{fg:msr}.

In our apparatus the polarimeter  is similar in general concept
to that used in  \cite{blatt}, except that the fringe magnetic field from the toroid is 
shimmed to produce an {\it azimuthal}  field at the muon stoppers. 
Muons are stopped in an annular array of aluminum stoppers 
surrounding the beam axis. Positrons from muon decay are detected in 
scintillator arrays located azimuthally between the muon stoppers so 
that an asymmetry between clockwise, $cw$- and counter-clockwise, $ccw$-decaying muons can be measured.
In our case, the 12-segment
polarimeter accepts muons exiting the gaps of the toroid and
is located well removed from both the stopping target and 
beam axis where background rates can be  reduced to a low level. This 
minimizes the problems of accidental rates in detecting the decay 
positrons and allows a long 
acceptance time for the muon decay ($\sim 20 \mu$s) in order obtain 
higher statistical sensitivity and detailed determination of the background.

One advantage of using the longitudinal field method in the present
experiment is higher statistical significance from a given number of 
positron events compared to a reduction of $1/\sqrt{2}$ in the case of 
the precession measurement. Although usually it is problematic how to remove 
instrumental misalignment which gives rise to a spurious asymmetry effect in 
the longitudinal field method, in the present experiment such 
effects can be canceled out by taking the ratio of $fwd$ and $bwd$ 
pions. 

$P_{T}$ would be indicated by an asymmetry between  
$cw$ and $ccw$ decay positrons as
\begin{equation}
\frac{N(cw)}{N(ccw)} \simeq 1 \pm 2 \alpha 
\langle\cos{\theta_{T}}\rangle P_{T},
\label{e1}
\end{equation}
where $\alpha$ is the effective analyzing power and 
$\langle\cos{\theta_{T}}\rangle$ is the kinematical attentuation relating 
the polarization axis to the decay plane.

Furthermore, it is possible to obtain the polarization from the double ratio
\begin{equation}
\frac{[N(cw)/N(ccw)]_{fwd}}{[N(cw)/N(ccw)]_{bwd}} \simeq 1 + 4 \alpha 
\langle\cos{\theta_{T}}\rangle P_{T},
\label{e2}
\end{equation}
with a doubling of the sensitivity to $P_{T}$. Several 
sources of potential systematic errors due to instrumental misalignments 
cancel out to first order in this double ratio. 
A direct 
measurement of the analyzing power can be obtained from the $P_{N}$ component, which is the in-plane component parallel to the pion momentum,
by selecting 
events with a transverse $\pi^{0}$ decay. The 
field requires good mapping and stability, but it does not need to be 
reversed as did the axial field in the previous experiment \cite{blatt}. 
This 
avoids potential systematic effects due to the reversing field itself as well as 
its effect on photomultiplier tube (PMT) gains.

\subsection{Overall assembly}
\label{subsec:assembly}

\begin{figure}
\epsfxsize=\linewidth
\begin{center}
\epsffile{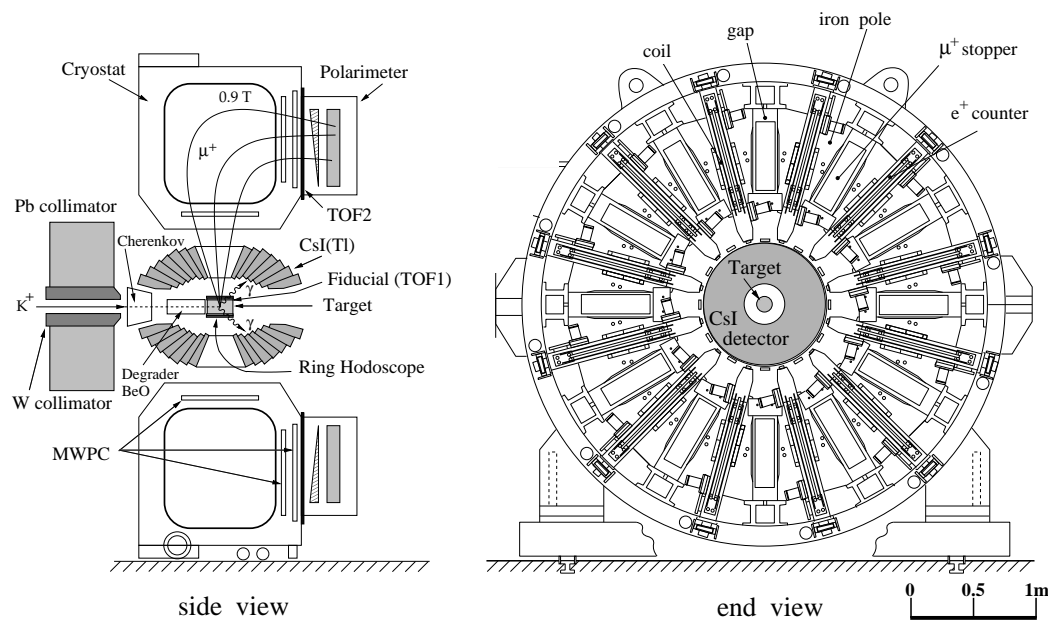}
\end{center}
\caption{General assembly side and end views of the detector.}
\label{fg:e246det}
\end{figure}

General assembly views of the apparatus are shown in  Figure \ref{fg:e246det}
as end and side views.
Our detector's three main analytical elements are the toroidal 
spectrometer, the CsI(Tl) photon calorimeter \cite{csinim}, 
and the polarimeter. 

The space available for the CsI(Tl) array is limited by the 1-m inner 
diameter of the toroidal magnet and cryostat. The crystal length of 
250 mm (13.5 radiation lengths) is adequate to keep shower leakage for photons up 
to 250 MeV acceptable while still allowing sufficient space for the 
PIN diode and preamplifier readout system. The crystal array has 12 holes aligned with the magnet gaps which provide the acceptance for charged particles in the spectrometer.

Two additional holes in the  crystal array were required on the beam 
axis: The upstream hole allowed for the incoming kaon beam and 
instrumentation which included a \v{C}erenkov detector, a scintillator 
hodoscope and a degrader. 
This hole had to accommodate not only the kaon beam 
phase space but also the six-times-higher pion contamination
including an asymmetric halo around the kaon beam, which could result 
in unacceptable rates in the crystal elements near the beam axis. This 
problem was partially alleviated by a Pb and W collimator upstream of 
the apparatus.
Downstream, the beam-axial hole provided access for the support and readout of 
the segmented stopping 
target assembly and inner tracking elements. The latter include the 
12 close-packed fiducial counters used in the event trigger and for 
muon time-of-flight measurements, and a stack of ring scintillators 
and a cylindrical drift chamber both used for muon tracking.

The target size was matched to the measured kaon beam 
profile at the stopping location at the center of the toroid. It is 
necessary to minimize the target size to avoid excess mass in the path 
of exiting muons and photons from $\pi^{0}$s, as well as to 
economize on the number of segmented elements in the target to be 
instrumented.

The polarimeter was mounted immediately downstream of the spectrometer 
to maximize acceptance for $K_{\mu3}$ $\mu^{+}$s in 
the 12 pure aluminum stopping arrays, and to take advantage of the shimmed 
fringe field of the toroid which serves as a holding field for 
$P_{T}$. A wedge shaped copper degrader preceded each aluminum muon stopping 
array, and compensated for the momentum dispersion from the analyzing 
magnet so that all muons would come to a near uniform stopping depth. In 
addition to the stoppers and positron detectors, the polarimeter 
included several veto counters used to help identify good stopping 
muons and valid decays while rejecting backgrounds.

\subsection{Acceptance optimization}
\label{subsec:acceptopt}

\begin{figure}
\epsfysize=\linewidth
\begin{center}
\epsffile{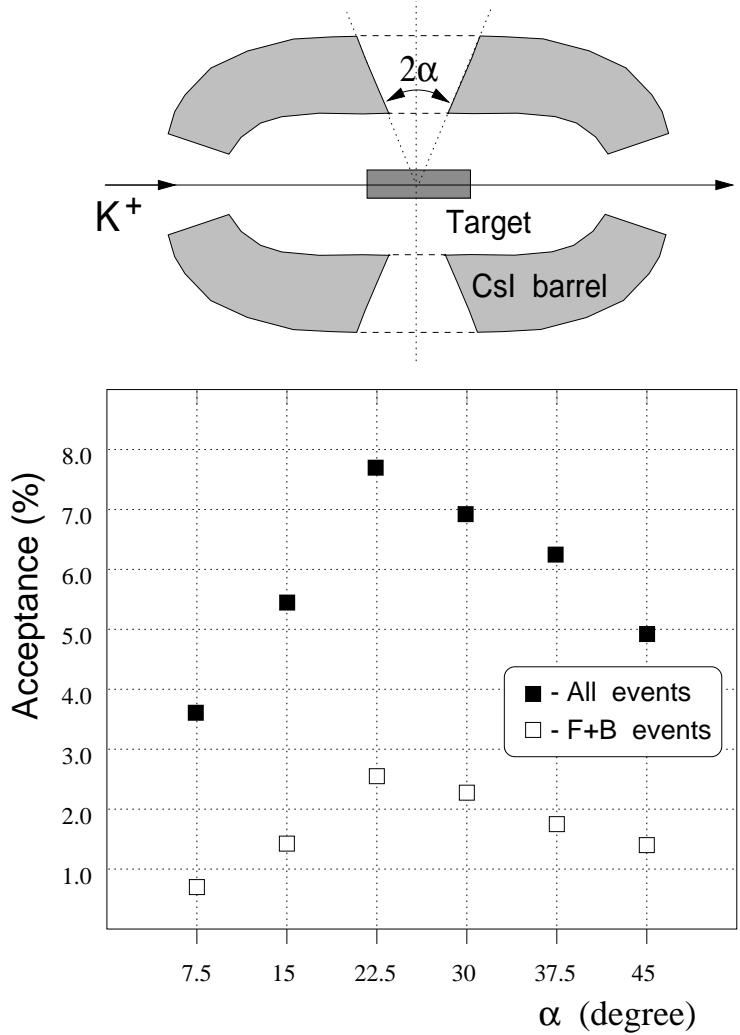}
\end{center}
\caption{Monte Carlo optimization of the muon holes. F+B refers to events with
forward- and backward-going pions.}
\label{fg:acceptance}
\end{figure}

The first design goal was to simultaneously
optimize the acceptance 
for $\mu^{+}$s from $K_{\mu3}$ in the 12 magnet gaps, and
to maximize the acceptance  
for $\pi^{0}$s in the CsI(Tl) calorimeter. This required careful 
matching of the 12 holes in the CsI(Tl) array to allow muons from the
distribution of stopped kaons in the 
central stopping target to enter the magnet gaps and be accepted in 
the polarimeter at the magnet exit. A Monte Carlo study 
\cite{csinim} lead to a design with  10\% acceptance for $\mu^{+}$s 
from $K_{\mu3}$, while maintaining 70\% coverage for $\gamma$ detection.
Optimum overall acceptance of 8\% was obtained for the angle $\alpha$ of 22.5 degrees as illustrated in Figure \ref{fg:acceptance}. 

\subsection{Systematic error reduction by symmetry}
\label{subsec:detsymmetry}

Our apparatus is constructed 
with 12-fold azimuthal symmetry  set by the 12-gap toroidal 
spectrometer magnet. The summation of events from the 12 spectrometer 
gaps contributes to the acceptance, and also allows additional 
capabilities to reduce systematic asymmetry effects due to the instrumentation 
and the kaon beam phase space. For example,  each positron detector in 
the polarimeter serves as a $cw$ counter for one gap and a $ccw$ 
counter for the neighboring gap, and an apparent asymmetry due to 
variations in counter efficiencies cancel. Similarly, asymmetries 
induced by an asymmetric kaon stopping distrubution also cancel in 
the azimuthal sum over 12 gaps. In this case also, since the decay 
vertex is determined event-by-event, residual higher order 
asymmetry can be tested for and corrected.

\section{Beamline and Beam Instrumentation}
\label{sec:beam}

\subsection{Low momentum $K^+$ beam line K5}
\label{subsec:k5}

\begin{figure}
\epsfxsize=\linewidth
\begin{center}
\epsffile{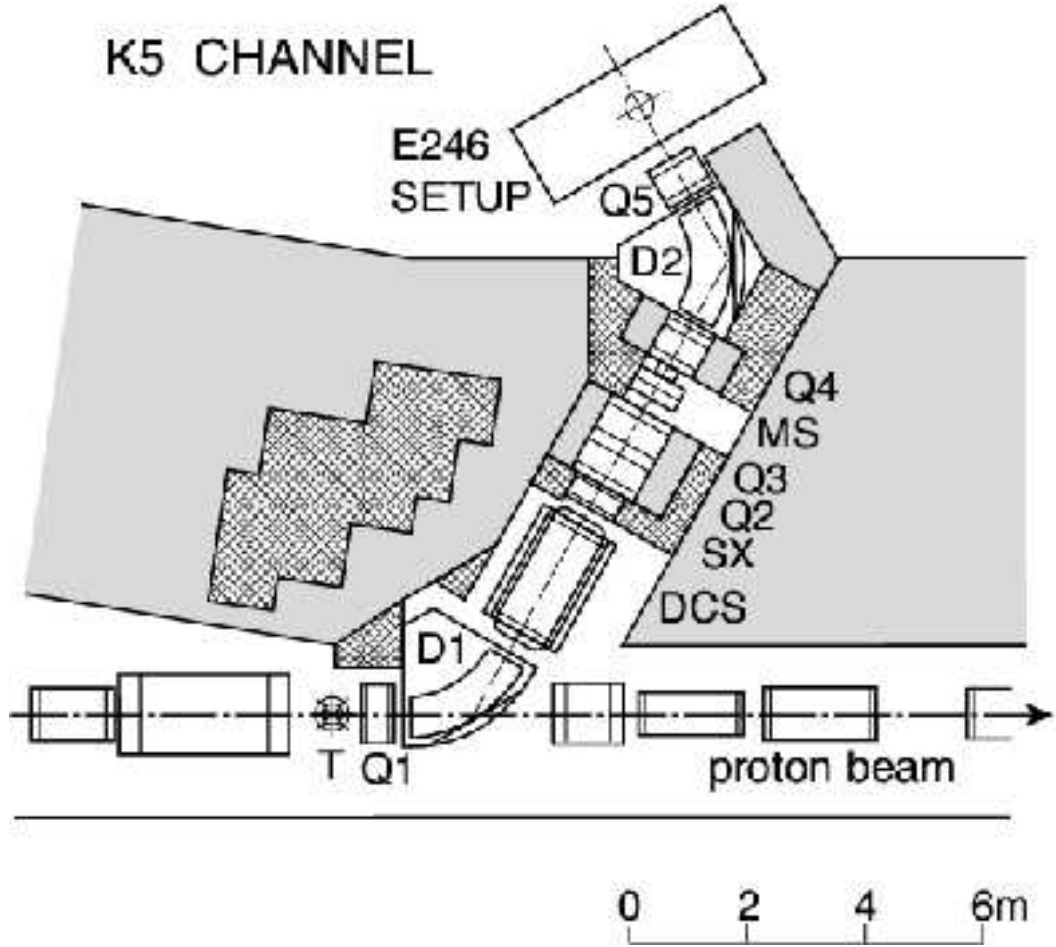}
\end{center}
\caption{Low momentum kaon beam channel K5.}
\label{fg:k5}
\end{figure}

The experiment was set up at the low-momentum separated kaon beam line K5 
\cite{k5bl} in the
North Hall of the KEK proton synchrotron. The layout of the beam line is shown
in Figure \ref{fg:k5} and the main parameters are listed in Table \ref{tb:k5beam}. 
This channel was originally designed for 550-MeV/$c$ maximum momentum,
 but it was upgraded to 660 MeV/$c$ for  this experiment. It is
equipped with a single-stage electrostatic separator (DCS) with an electric
field of 50 kV-cm$^{-1}$ and can deliver a $K^+$  beam with $\pi$:$K$ ratio of 
5 -- 10 making a nearly achromatic spot on the experimental target. 

\begin{center}  
\begin{table}
 \caption{Main parameters of the K5 beam channel} 
 \label{tb:k5beam}
 \begin{tabular}{l l}
   \hline
   maximum momentum       & 660 MeV                                \\
   production target      & 6(H)$\times$10(W)$\times$60(L)mm$^{3}$     \\
   take-off angle         & $0\pm3$ degrees                               \\
   solid angle            & 10 msr ($\pm180$mr(H)$\times\pm18$mr(V))\\
   channel length         & 12.5 m                                 \\
   momentum bite          & $\pm3\%$                               \\
   beam size on target    & $\sigma_x = \sigma_y = 2.5$ cm         \\
   dispersion on target   & 0.28 cm/\%$\Delta p/p$                 \\
   $K^+$ beam intensity   & $1.0 \times 10^{5} / 10^{12}$ protons    \\
   $\pi^{+}/K^{+}$ ratio  & $ 6 \sim 7$                                     \\   
   \hline
 \end{tabular}
\end{table}
\end{center}

Secondary particles were produced at a 6(H) $\times$ 10(W) $\times$ 
60(L) 
mm$^3$ Pt target in the 12-GeV proton beam with a typical intensity of 
$3\times 10^{12}$ protons per pulse of 0.7-s duty with 3.0-s repetition. 
The secondary beam was collected at 0 degrees with a large solid angle of
$\sim 10$ msr. Two bends D1 and D2 and a horizontal acceptance slit at the 
intermediate point selected the $\pm 3\%$ beam momentum bite. The slit 
serves also to reduce the intensity if necessary. The mass separation 
is done after the DCS at the  vertical mass slit after the 
intermediate quadrupole 
doublet of Q2 and Q3. The last doublet of Q4 and Q5 focuses the beam on the 
target with a round spot. The momentum dispersion at the target is 
0.28 cm-$(\%\Delta p/p)^{-1}$.  The total length of the channel is 12.5 m. 

In a Monte Carlo simulation it was shown that the optimum 
momentum to obtain the maximum kaon stopping rate in the fiducial region of 
our target was 700-800 MeV/$c$ taking into account the momentum dependence of 
the kaon production cross section and scattering of the beam from the degrader. 
Accordingly, the  maximum channel momentum of 660 MeV/$c$ was chosen for the 
experiment. 

Beam tuning was done to maximize both the number of stopped kaons and 
the $K$:$\pi$ ratio as well as to realize a symmetric kaon 
stopping distribution in the target. Since the channel has only one stage 
of DCS, it was difficult to reduce pion contamination which arises 
from $K_s$ decay near the proton target. In addition,
a pion beam halo 
made background hits in the CsI(Tl) calorimeter. In order to remove this 
halo a ray-tracing study was performed for kaons and pions identified
separately by a \v{C}erenkov counter and back-tracking the beam by means of 
a MWPC at the target position and a horizontal counter hodoscope placed after 
the mass slit. Rejection of the pion halo was realized by installing a beam 
slit inside the second bend D2. The typical beam intensity at 660 MeV/$c$ of $K^+$ is
$10^5$ per $10^{12}$ proton with a $\pi^+/K^+$ ratio of 6 -- 7.

\subsection{Beam collimation and degrader}
\label{subsec:colldeg}

Between the last quadrupole of the K5 channel and the beam \v{C}erenkov
counter
a collimator system was inserted. It consists of a 45-cm thick lead wall with
a rectangular hole with an additional insert made of  Hevimet  
for shaping the
beam profile. The thickness was determined so as to stop the 
660-MeV/$c$ halo pions. The shape of the insert was designed
to transmit the kaon profile while also blocking 
the pion halo in the CsI(Tl) region.
This collimator system cut the pion halo background by 70\% while
sacrificing only 15\% of the kaon yield, and reduced the singles counting rate
of the affected CsI(Tl) crystals drastically.

After passing through the \v{C}erenkov counter the beam was slowed 
by a degrader system consisting of 65 mm of Al followed by 242 mm
of BeO located close to the stopping target. The 100-mm diameter 
was enough to cover the beam profile. The use of light-element 
materials is essential to achieve high stopping power with
minimum multiple scattering; in addition, using BeO near 
the target helped to reduce the interaction of photons from the target 
in the degrader while the use of Al further upstream served to 
reduce the cost. The combination of lengths was optimized for a 
stopping distribution centered in the 20-cm long fiducial region of the 
target. The broadening of the beam through the degrader was not 
negligible but acceptable.

\subsection{Beam detectors}
\label{subsec:beamdet}

\subsubsection{\v{C}erenkov}
\label{subsubsec:cerenkov}

\begin{figure}
\epsfxsize=\linewidth
\begin{center}
\epsffile{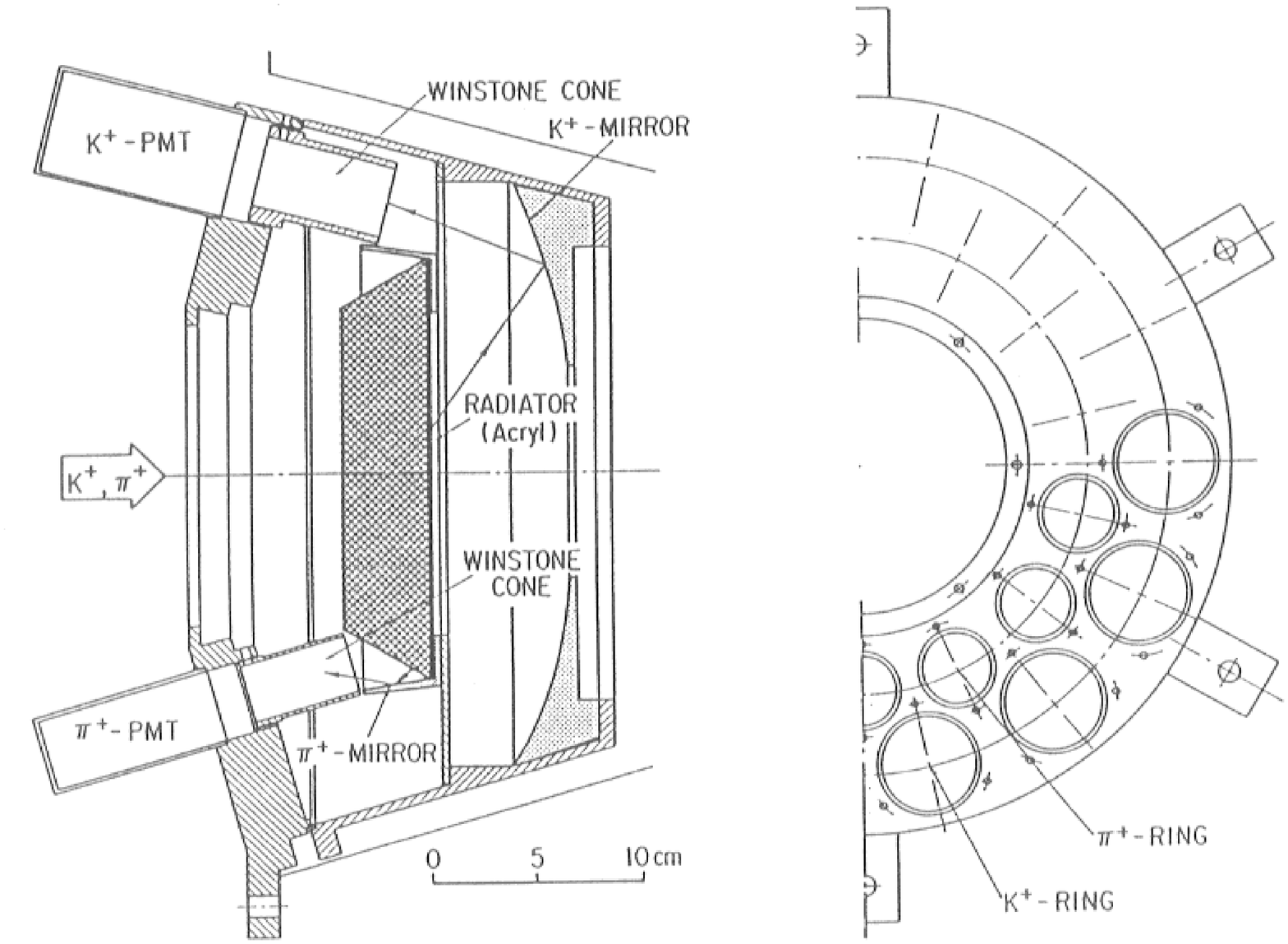}
\end{center}
 \caption{Cross section side view and end view of the beam 
 \v{C}erenkov counter.}
 \label{fg:cerenkov}
\end{figure}

In order to trigger on kaons in the beam, a
Fitch type differential \v{C}erenkov counter was employed \cite{cerenkov}. 
Figure \ref{fg:cerenkov} shows the crossection side view and end view of
the counter.  Incident pions and kaons emit \v{C}erenkov 
light in a 4-cm thick
acrylic radiator with a velocity dependent emission angle. The index 
of refraction of 
 the radiator is such that at our chosen momentum
the kaon light emerges from the rear
surface while the pion light is internally reflected and transmitted 
to emerge from the circumference of the radiator.  Two circular 
arrays of
 PMTs are arranged so that the inner circle detects the
pion light from the radiator circumference, and the outer circle
detects the kaon light which is reflected by a curved mirror behind
the radiator.  Each array comprises 14 PMTs with a Winston cone in
front.

The configuration of the radiator, mirrors and PMTs was optimized
to maximize the total number of PMT 
sensitive photons for incident
beam momenta in the range 620 -- 700 MeV/$c$.  Additionally, the
design was constrained to minimize the thickness of the radiator to
avoid excessive multiple scattering of the beam, to optimize its effective aperture
to cover the whole beam, and to minimize overall dimensions in order
to fit inside the entrance aperture of the CsI(Tl) array.

The number of PMTs was determined by the available room. 
The
parabolic kaon mirror and kaon Winston cones were made of acrylic
with a reflective coating of
evaporated Al.  The pion mirror was made of aluminized Mylar and
the pion ring Winston cones were coated with white BaSO$_{4}$.  Preceding the
experiment the basic characteristics of the counter were tested at 
 the KEK 12-GeV synchrotron using a pion beam of the
same velocity as the pions and kaons in the experiment.  The effective
radius with more than 90\% detection efficiency was measured to be 6
cm.  The main parameters of the counter 
are summarized in Table \ref{tab:che}.

\begin{center}   
\begin{table}
\caption{Main parameters of the beam \v{C}erenkov counter}
\label{tab:che}
\begin{tabular}{ll}
\hline
kaon momentum range          & 620 to 700 MeV/c     \\
radiator material            & acrylic  ($n_D = 1.49$)     \\
radiator thickness           & 4.0 cm                \\
effective aperture           & 12-cm diameter       \\
beam angular tolerance       & $\pm 3^o $ for $r < 4$ cm   \\
kaon ring                    &                      \\
~~PMTs                &Hamamatsu R580UV $\times$ 14    \\
~~Winston cone              & 5.3 cm $\phi$ with Al coating \\    
pion ring                    &                      \\
~~PMTs                &Hamamatsu R1398 $\times$ 14     \\
~~Winston cone              & 3.8 cm in diameter with BaSO$_{4}$ coating \\
\hline
\end{tabular}
 \end{table}
\end{center} 

\begin{figure}
\epsfxsize=\linewidth
\begin{center}
\epsffile{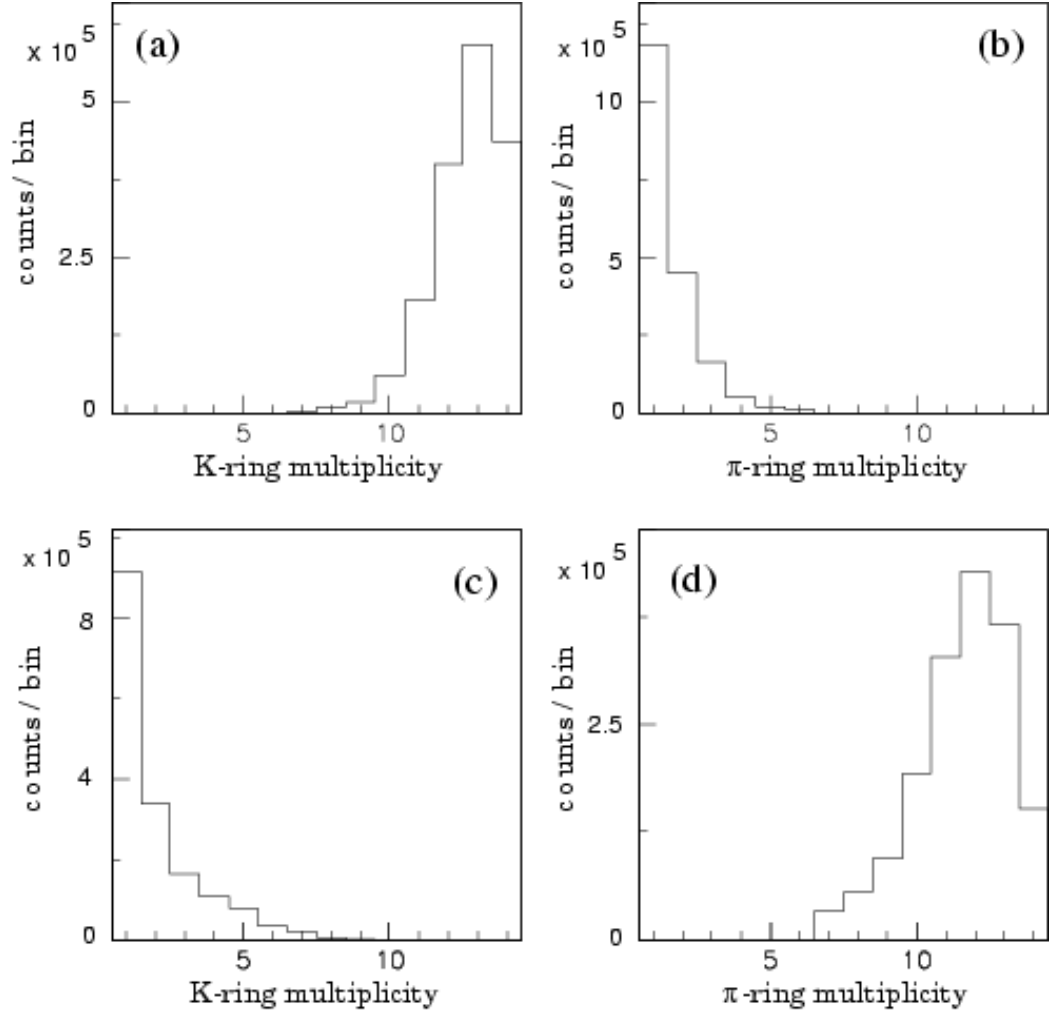}
\end{center}
\caption{Multiplicity spectra of the kaon ring (a ,c) and pion ring (b,d). 
The upper spectra are for kaon incident and the lower spectra are pion
incident.  Particle identification was done by
TOF.
}
\label{fg:che2}
\end{figure}

For the online trigger, the multiplicity of each PMT ring was used.  The
trigger condition was investigated in a test experiment by identifying
particles using time-of-flight.  Figure \ref{fg:che2} shows the typical
multiplicity spectra of both rings for both kaons and pions incident. 
In  normal operation the kaon trigger threshold was set to
multiplicity of $\ge 6$ with a trigger efficiency of more than 99\%.  The
probability of mis-identification of a pion as a kaon was less than 1\%
under the typical beam intensity conditions. 

\subsubsection{Hodoscope}
\label{subsubsec:hodoscope}

\begin{figure}
\epsfxsize=\linewidth
\begin{center}
\epsffile{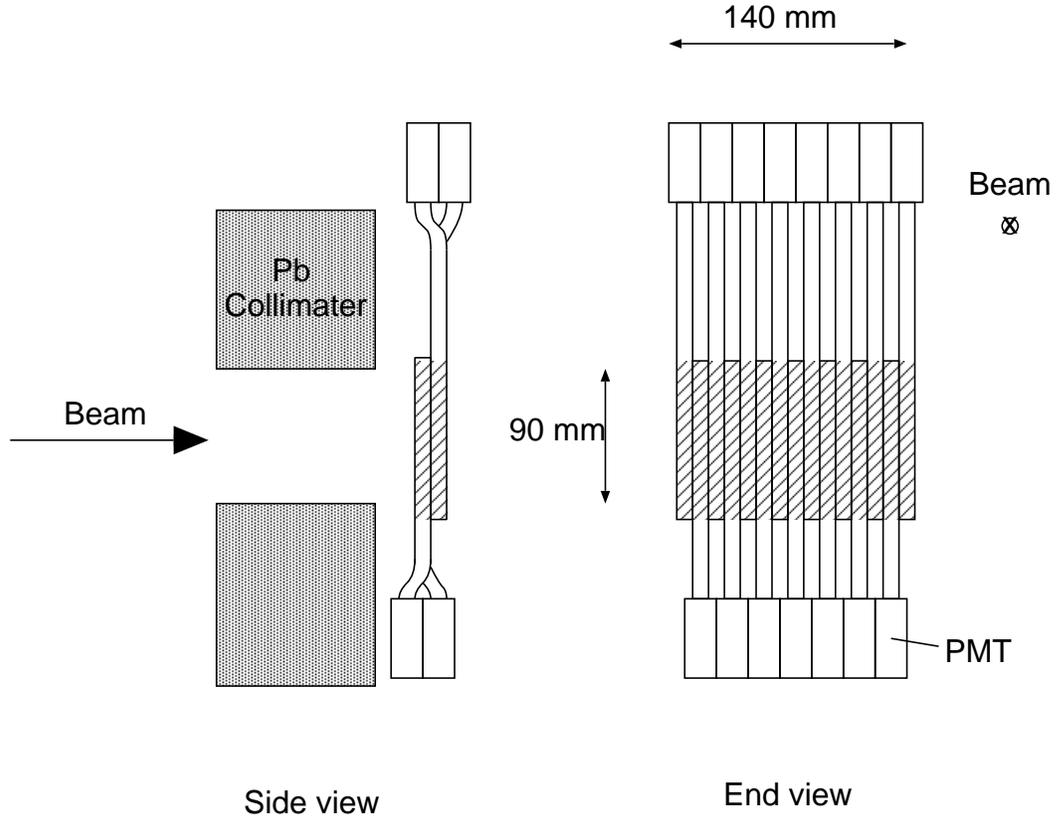}
\end{center}
\caption{Schematic diagram of the B0 beam counter and its location relative to the collimator.}
\label{fg:b0}
\end{figure}

The B0 hodoscope 
shown in Figure \ref{fg:b0} was placed downstream of the Pb collimator, and 
is an assembly of 22 plastic scintillating counters. $K^+$ and $\pi^{+}$ beam
profiles were obtained in coincidence with the \v{C}erenkov
counter. The beam profile was monitored to check the stability of 
both the
primary proton beam and the secondary particle beam. The trajectories of
beam particles were also obtained by using the $K^+$ entrance
position determined by the active target. 

Timing data of
charged particle hits in the B0 hodoscope were recorded by multi-hit 
TDCs
during the $e^+$ gate for the $\mu^+$ decay.  Hits in the polarimeter positron
counters in time with B0 formed part of the constant background in the $\mu^+$
decay spectra related to beam particles and were rejected by vetoing 
on
coincident events in the B0 hodoscope.

\section{Charged Particle Measurement}
\label{sec:chpartmeas}

\subsection{Component overview}
\label{subsec:chpart overvw}

Several detector 
elements were used to measure charged particles from the decay of stopped kaons. The main functional goals of this system are particle identification and kinematic reconstruction using time-of-flight (TOF) and tracking through the magnetic spectrometer before entering the polarimeter.

In the central region, the segmented active target  gave the lateral coordinates ($x,y$) for the stopping position and decay vertex of the kaon as well as the initial track of exiting charged particles. The fiducial counters surrounding the target provided the initial trigger information and the start for the TOF measurement. Immediately outside the fiducial counters, the ring hodoscope gave the first axial ($z$) track coordinate before passing through the gaps in the CsI(Tl) calorimeter to enter the toroidal spectrometer. A cylindrical drift drift chamber (C1) just outside the ring hodoscope gave additional tracking information. At the entrance and exit of each spectrometer gap were sets of planar wire chambers (C2, C3, C4) which gave the necessary coordinates for momentum analysis. Finally, the charged particles entered the polarimeter by passing first through the second time-of-flight counter (TOF2) and then a defining polarimeter trigger counter (PL).

\subsection{Target and ring counters}
\label{subsec:tgtrings}

\subsubsection{Active target}
\label{acttgt}

A segmented active target identified the stopping kaon and its decay 
vertex, and provided the initial tracking information  for the decay 
muon as well as an energy-loss correction to the momentum measurement 
in the spectrometer. The segmentation level is a trade-off between 
position resolution and rate handling 
from fine segmentation on the one hand, and energy 
and timing resolution from minimum ionizing decay muons and lower 
cost from coarse segmentation on the other hand. Based on studies of 
scintillating fibers of various sizes as well as from the experience 
with the target used in the E787 experiment at Brookhaven 
\cite{e787target}, we chose an array of $5 \times 5$ mm$^{2}$ Bicron 
BCF-12 fibers. 

\begin{figure}
\epsfxsize=\linewidth
\begin{center}
\epsffile{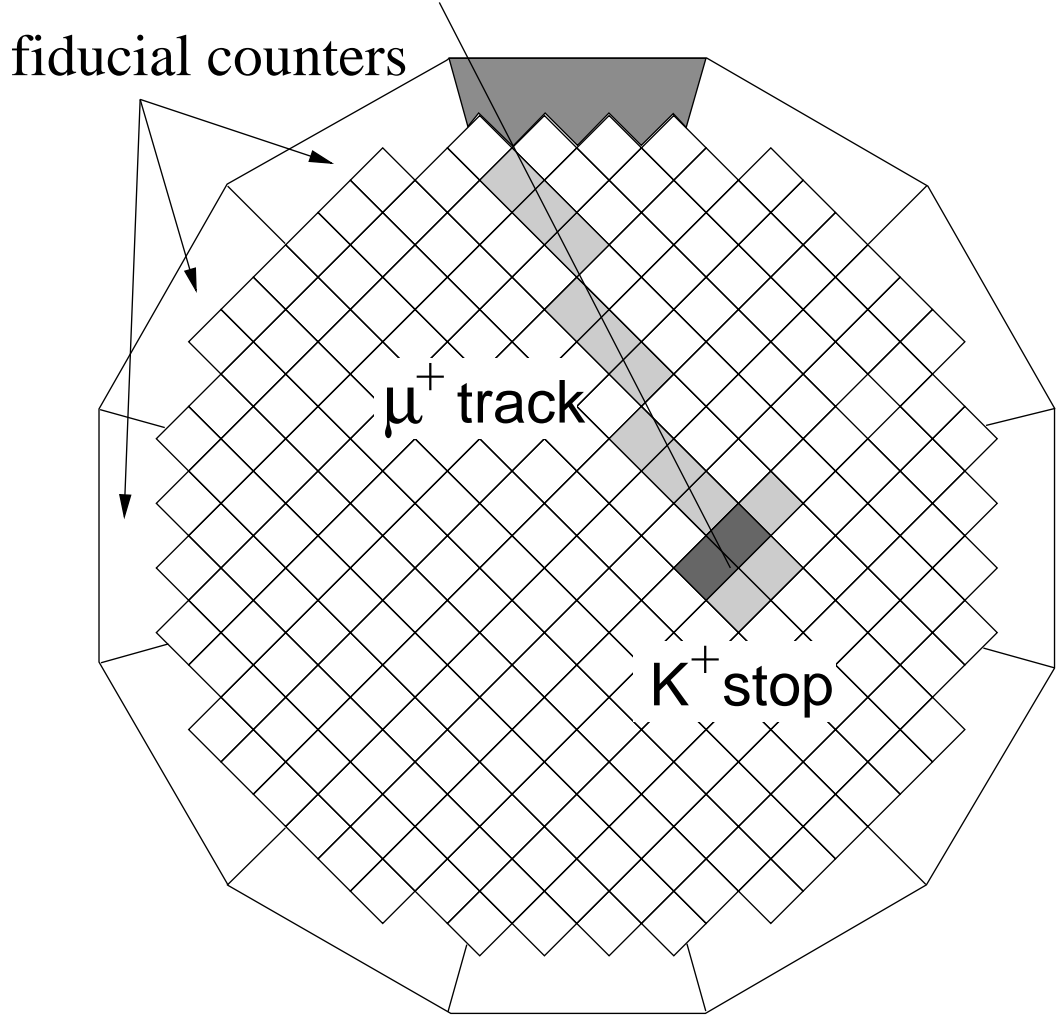}
\end{center}
\caption{Crossectional view of the active target taken from the online event display, showing the elements hit by the stopping kaon and the exiting charged particle (muon). The fiducial counter hit contributing to the event trigger is shown at the top. The shading roughly indicates the energy loss in each element.}
\label{fg:tgtxsec}
\end{figure}

 The radial extent of 
the target was optimized to match the kaon beam profile, while minimizing 
the material to be traversed by exiting decay muons and photons 
from $\pi^{0}$s. The optimum was a $\sim$9-cm diameter array of 256 
fibers shown in Figure \ref{fg:tgtxsec}. The 45$^{\circ}$-diagonal 
orientation was chosen to maximize the vertical moment of inertia of 
the individual fibers to resist sagging under gravity, since the array 
is cantilevered from its support downstream of the active center 
region of the detector.

In order to determine the target stopping efficiency with respect
to incident kaons several measurements with a special trigger
consisting only of the kaon \v{C}erenkov counter were carried out. 
The efficiency was 38\% for kaons identified in the target 
as events with a large energy deposit
in one fibre ($E_{fiber} > 5$ MeV) normalized by the number of
triggers.  The same efficiency was obtained when the 3-cluster events
in the CsI(Tl) detector were taken into account.
The stopping efficiency of 38\% was consistent with preliminary estimations
($\sim 40\%$) based on the beam profile and target dimensions.

\begin{figure}
\epsfxsize=\linewidth
\begin{center}
\epsffile{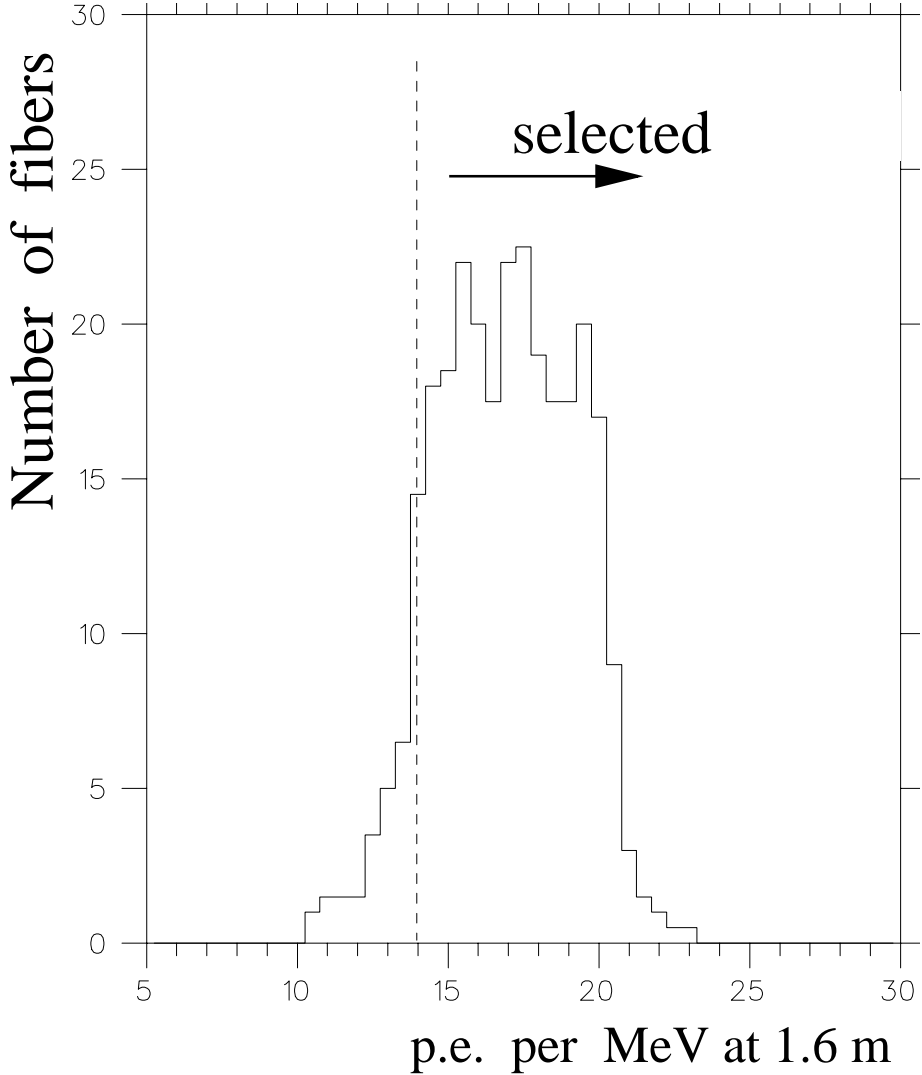}
\end{center}
\caption{The distribution of light yields in photo-electrons per MeV of fibers tested for the target, showing the criterion for the 256 fibers selected.}
\label{fg:pemev}
\end{figure}

After selection for crossection dimensions and lack of ``pin cushion''
distortion, one end of each 2-m long BCF-12 fiber received from Bicron 
was machined with a diamond cutter to be 
coupled to a PMT in a test setup where scintillation light 
output and attenuation length were measured. 
The 
measured attenuation lengths of most fibers fell in a range 200 -- 400 
cm, however the parameter of practical concern is the light output 
in photo-electrons per MeV (p.e./MeV) measured at a fixed distance of 1.6 m from 
the PMT, which approximated the conditions in the actual target. The 
distribution is shown in Figure \ref{fg:pemev} for the 282 fibers 
measured. Fibers with a minimum light yield of 14 p.e./MeV were 
accepted for the target with an average for the 256 target fibers of 17.3 
p.e./MeV. It was found that changing the PMT coupling from an 
air gap to a silicone gel coupling increased the light output by 12\%.  
Nevertheless, since the light output was adequate with an air gap, 
this simpler arrangement was adopted in the target assembly.

The selected fibers were then diamond machined to the finished length 
of 185 cm, and spiral wrapped with 6.4-$\mu$m aluminized mylar. The 
main purpose of the mylar wrapping was to protect the fibers in the 
target array and to avoid the possibility of light cross talk between 
fibers. Wrapping with white teflon tape was also investigated but neither 
material produced a significant improvement in the light 
output. This can be largely attributed to the air gap coupling to the 
PMT which preselects light rays with near normal exit angles and 
thus shallow angles of reflection propogating in the fiber, where the 
dominant effect is internal reflection at the fiber cladding and not 
the outer surface.

Figure \ref{fg:tgtxsec} also shows the 
close-packed  fiducial counters surrounding the fiber array.
These counters will be discussed further in Section \ref{subsec:tofctrs}. The 
array of target fibers and fiducial scintillators/lightguides was 
assembled in a jig and the active end was 
wrapped to form a rigid light tight ``log''. The position of each 
fiber in the array was documented so that balancing of 
higher-gain PMTs with lower-light-yield fibers was possible. The ``log'' 
was then clamped in a support collar from which the active end of the 
target array was cantilevered in the final assembly. The opposite 
ends of the fibers were free to fan out to individual PMTs (Hamamatsu 
H3164PX MOD).

Two support assemblies are attached to the support collar. One 
extends below to a base which slides on a track allowing the target to be 
positioned and aligned in the detector and easily moved for access to other 
systems. The other support consists of six struts 
which hold the PMTs mounted on a facetted plate approximating a section of 
a spherical surface, where the fibers are fanned out to the PMTs.
The facetted plate is also mounted on the sliding base.

The 256 PMT signals from the target fibers were fed to custom made 
postamplifier-discriminator modules housed in a TKO crate. The postamplifiers
 had a ten-fold gain output delayed by
a built-in hybrid circuit for 100 ns. The analog output fed the ADC and
the output of an incorporated leading-edge discriminator was used for the TDC.

Figure \ref{fg:tgtxsec} illustrates the on-line display of a typical 
$K_{\mu3}$ event showing the cluster of fibers hit by the stopping 
kaon, and the track from the exiting decay muon. 
The $K^{+}$ decay vertex in the plane perpendicular
to the beam axis is obtained from the target
 scintillating fibers with an accuracy of 
$\sigma$~=~1.4~mm. The main parameters of the target are summarized in Table \ref{tb:tgtrings}.

\begin{center}
\begin{table}
 \caption{ Main parameers of the target array and ring counters.}
 \label{tb:tgtrings}
 \begin{tabular}{ll}\hline
Target parameters& \\
~~number of elements (fibers) &  256   \\
~~scintillator type   &   Bicron BCF12   \\
~~fiber dimensions   &  $5 \times 5 \times 1850 $ mm$^3$  \\
~~array ``diameter''   &  $93 \pm 3$ mm  \\
~~light yield (av.)  &   $17.3 \pm 4 $ p.e./MeV (MIP) \\
~~PMTs    &    Hamamatsu H3164 PX MOD \\
~~PMT -- fiber coupling   &   air gap  \\
Ring parameters & \\
~~number of elements (rings)  &    32   \\
~~scintillator type  &   Bicron BC408   \\
~~WLS type   &   Kurraray Y11 ( 1-mm diam.)  \\
~~ring dimensions  &  ID 118 mm, OD 128 mm, $z$-thickness 6 mm   \\
~~light yield (typ.)  &    46 p.e./MeV (MIP)  \\
~~PMTs   &   Hamamatsu R580-17   \\
~~PMT -- WLS coupling   &  optical grease \\
\hline
\end{tabular}
\end{table}
\end{center}

\subsubsection{Ring counters}
\label{subsubsec:ringctrs}

The coordinate along the beam axis ($z$-coordinate) was recovered with 
the hodoscope  surrounding the target and fiducial counter 
assembly. The design is a cylindrical 
array of
plastic scintillator rings employing wavelength-shifting fiber readout \cite{ringNIM}.
The  hodoscope 
parameters were optimized to obtain 
high efficiency for detection of minimum ionizing particles 
(MIPs) in the
high rate environment, and to achieve better than 2-mm spatial resolution.  
The main parameters are summarized in Table \ref{tb:tgtrings}.

\begin{figure}
\epsfxsize=\linewidth
\begin{center}
\epsffile{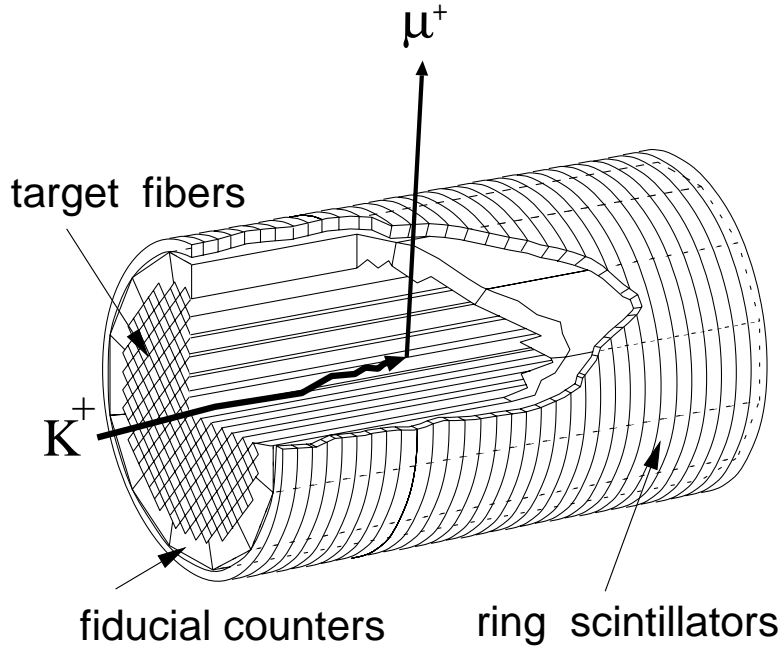}
\end{center}
\caption{Isometric view showing the 32 ring scintillators surrounding the active target and fiducial counters. The WLS fibers, not shown, were bent and exited to the right to the PMTs located downstream.}
\label{fg:rings}
\end{figure}

A schematic view of the ring system is shown in 
Figure \ref{fg:rings}. The rings of 118-mm inner and 128-mm 
outer  diameters were cut
from 6-mm thick BC408 scintillator sheet
using a diamond cutter.  The 5-mm radial ring thickness was  
a compromise between high
light output and minimal additional material 
causing particle scattering and photon conversions. 

Scintillation light is trapped inside 
the ring
by using a highly reflective wrapping until it is captured by the WLS fiber. 
Since the
light is susceptible to inevitable losses before entering the fiber, 
the optimization requires high 
reflectivity at the scintillator-wrapping boundary,
high transmittance at the scintillator-fiber boundary, and
large geometrical acceptance of the fiber.
The first factor is dependent on the wrapping and scintillator surface 
treatment, while
the second is determined by the optical couplant, the surface 
quality in the contact
area and the ratio of  refractive indices between the scintillator 
and fiber.

The  spectral matching of 
multi-clad Kuraray Y11  WLS fiber in combination 
with  Bicron BC408 scintillator
provides  high light yield \cite{y11-bc408}. 
 The Y11(K27) emission peak of 495~nm is 
shifted to 510~nm after
the light travels 1.5~m through the fiber to the PMT. 
 A fiber with a diameter of 
1.0 mm
was chosen as a reasonable compromise between light output and fiber 
flexibility.
To increase the geometrical acceptance, the
 fibers are placed in spiral
 grooves machined into the outer circumference of
 the  scintillator rings and glued using a 
colorless silicone adhesive SE777\footnote{Toray Dow Corning Silicone, Japan.}. 

The two $\sim$1.3-m long 
ends of the fiber were inserted 
into a light-tight 
flexible tube and the tips were glued into an acrylic holder. 
After gluing the fiber into the holder its ends were cut off 
and  polished.
The holder
provides a mechanical fiber-phototube interface and accurately 
controls
the perpedicular alignment between the fiber axis and the PMT window.
The supporting acrylic also prevents any erosion or chipping of the
fiber edge.

The 32 ring-fiber assemblies were positioned on a support cylinder made 
of
0.5-mm thick G10 sheet. The rings are interleaved with aluminized mylar
to prevent  light crosstalk. The wrapping material  causes a 0.30-mm
gap between neighboring rings which resulted in a dead fraction of 4\%
for the hodoscope. The hodoscope outer diameter is limited
by the available space requiring that the fibers be bent and 
held with  adhesive tape to
the detector surface using a minimum 2-cm
bending radius to route the fibers downstream of
the target to the PMTs. The assembled detector  
was shielded by a protective outer cylinder wrapped with  black
light tight tape. After installing the hodoscope and fiber readout 
``tails'' onto the
target, the acrylic holders were attached to the Hamamatsu R580-17 
phototubes using  high-viscosity optical grease.

\subsection{Superconducting spectrometer magnet}
\label{subsec:magnet}

\begin{figure}
\epsfxsize=\linewidth
\begin{center}
\epsffile{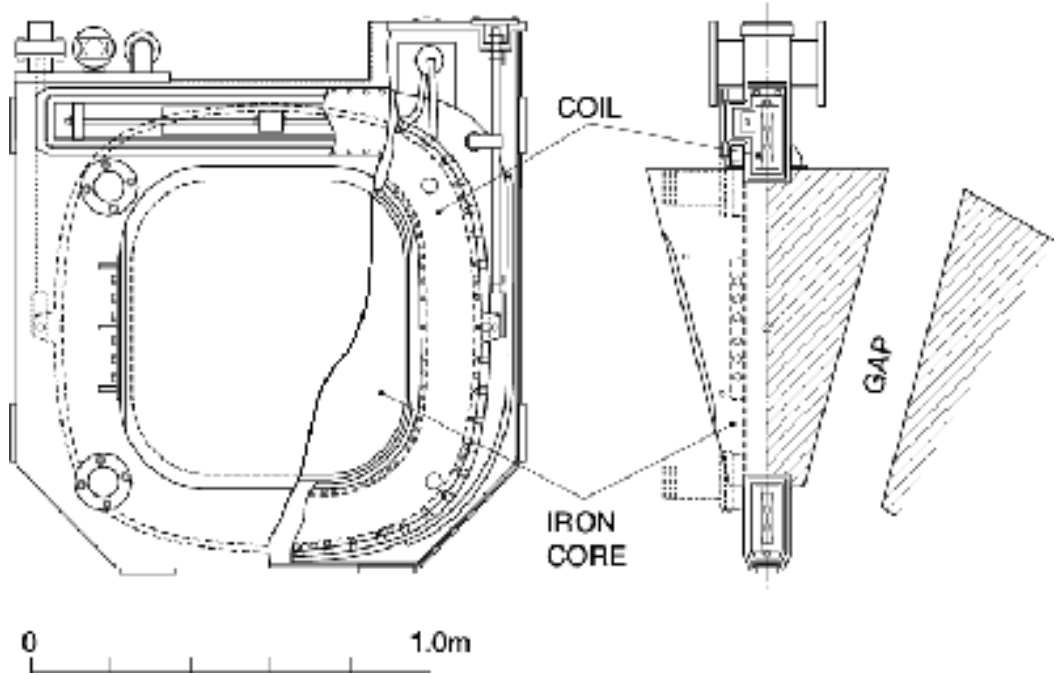}
\end{center}
\caption{One sector of the superconducting toroidal magnet.}
\label{fg:magsector}
\end{figure}

The heart of the muon spectrometer is a superconducting toroidal 
magnet with 12 iron sectors separated by 12 gaps \cite{toroid}. Each iron sector is 
magnetized by a superconducting coil, and a field up to 1.8 T can be 
excited across the 20-cm uniform gaps. At a field strength of 0.9 T 
used for this experiment the field is nearly completely dipole with a slight 
toroidal component superposition. Charged particles from the target 
located in the center of the magnet are bent by $\sim90$ degrees and 
tracked 
by MWPCs at the entrance and exit of the gap. One sector of the 
magnet is illustrated in Figure \ref{fg:magsector}.

In manufacturing the magnet special care was taken to assure the 
dimensional accuracy necessary for high precision experiments. Each iron
core was machined with a precision of 
50 $\mu$m. The positioning of the superconducting coil relative to the core
to form a sector was achieved
within an accuracy of 2 mm. Assembly of the entire structure was 
performed using a special positioning device ensuring a rotational 
symmetry of 30 degrees. The 12 median planes of the gaps converge to a
virtual central axis with an accuracy of 0.3 mm. The difference of the
diameters in the horizontal and vertical directions is less than 2 mm
over a diameter of about 4.0 m.

As a superconducting magnet this device is unique in its
structure. The coil windings are NbTi monolith stabilized by Cu operated
at about half of the critical current density at 4.5 K at the maximum 
excitation. The windings are cooled indirectly by two-phase He 
flow through one turn of cooling channel around the windings. 

The twelve coils are connected in series and equal field distributions are
generated in the 12 gaps. A power supply with stability of $5\times 
10^{-5}$ is used. Because high quality magnet steel was used for 
the cores, there is no significant hysteresis in magnetization. 
Nevertheless, 
a definite excitation cycle from 0 T up to 0.9 T and from 0.9 T down to 0 
has been followed for the E246 experiment  in 
order to assure the reproducibility of the field strength.

The field map necessary for tracking was calculated by using a 3-dimensional
code TOSCA\cite{tosca}. The validity of the map could be checked from
the measured monochromatic momentum spectra of muons from $K_{\mu2}$ 
and pions from $K_{\pi2}$.  

\subsection{Tracking chambers}
\label{subsec:chambers}

Charged-particle tracking through the spectrometer is crucial 
in rejecting other kaon decay modes such as
$K_{\pi 2}$ and $K^+\rightarrow\pi^+\pi^{+/0}\pi^{-/0}$ ($K_{\pi 3}$) 
by momentum measurement, and in fixing
the kinematics of $K_{\mu 3}$ for the determination of the decay
plane.

To meet these objectives the tracking chambers needed to have 
position resolution better than 500 $\mu$m in sigma in
the direction parallel to the magnet pole. 
They must also be able to
measure the position of particles with steep incident angles (maximum
40$^{\circ}$), and have the
 capability to operate in the non-uniform magnetic fringe
field of the toroid up to 4 kG.

\subsubsection{C1}

Before the ring hodoscope was installed, the first tracking element outside the active target/fiducial counters was an annular cylindrical drift chamber, C1. It consisted of  four layers of 
drift cells with a central anode wire in each, between cylindrical cathode planes at the inner and outer radii of the annular cylinder.
The anode signals from the drift cells provided additional ($x,y$) coordinates outside the target, and were also able to detect photon conversion in the outermost layers of the target.

The cathode planes were etched to form $45^\circ$ spiral strips, with the inner and outer planes of opposite pitch, so that a measure of the axial track coordinate $z$, was obtained from the induced charge distribution from the inner-  and outermost drift cell layers.
Under the conditions of high intensity beam operation,
 the high rate experienced so near the stopping target proved  
the ring counter hodoscope  (Section \ref{subsubsec:ringctrs})  to be more reliable in extracting the $z$ coordinate rather than the C1 cathodes.

\subsubsection{C2, C3 and C4}

Sets of three planar multiwire proportional chambers (MWPC) were
installed at each magnet gap.  One MWPC was installed at the entrance
of the magnet gap (C2) and two were installed at the exit (C3 and C4).  

These MWPCs utilize cathode readout
in order to save readout space around the chamber and to offer high
position resolution in the non-uniform magnetic field.
Each chamber has two sets of cathode planes: one for the azimuthal (or
$y$) coordinate and the other for the polar (or radial) coordinate. 
The 20 $\mu$m-diameter Au-coated W anode wires run with 2-mm 
spacing parallel to the magnet gap median
plane in order to provide higher position resolution for the
cathode plane measuring the 
most sensitive coordinate 
for the momentum reconstruction.  The cathode strips running perpendicular
to the anode wires, are 9-mm wide, spaced 1-mm apart and 
consist of an 18-$\mu$m thick
Cu coating on Kapton foil. 
The half-gap between the anode and the cathode planes is 6 mm.  

\begin{figure}
\epsfxsize=\linewidth
\begin{center}
\epsffile{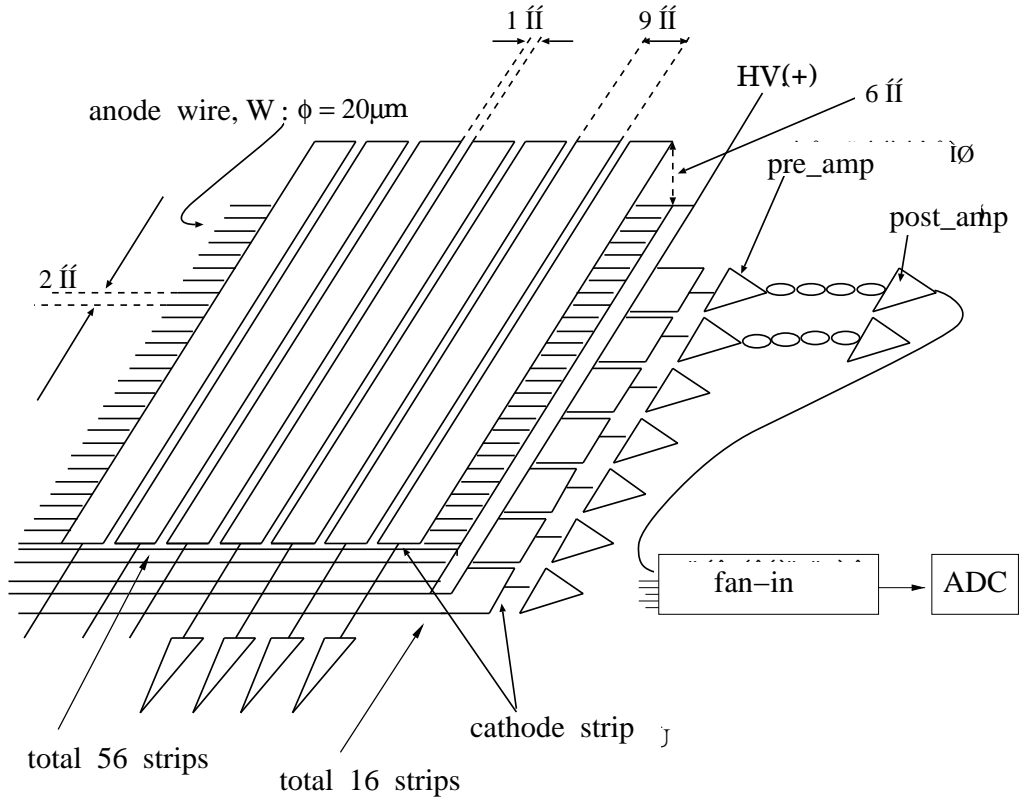}
\end{center}
\caption{Schematic view of the MWPC construction showing the cathode strip orientation with respect to the anode wires.}
\label{fg:view_mwpc}
\end{figure}

A schematic view of the MWPC
is shown in Figure \ref{fg:view_mwpc}. Given the spatial
constraints  near the gap, one important feature 
(in partuicular for C2) is the thin Al frame which maximizes the
sensitive area in the space available. 

The chambers were operated in proportional mode with a 50:50 Ar:ethane
gas mixture.  The three chambers in each gap are
connected in series with a gas flow of 30 cc-min$^{-1}$  from the gas
circulating system. 
The main parameters are summarized in Table \ref{tb:spec_mwpc}.

\begin{center}
\begin{table}
\caption{Main parameters of the planar MWPCs.}
\label{tb:spec_mwpc}
\begin{tabular} {l l }
\hline
anode wires       &     20 $\mu$-diameter Au-coated W  \\
anode pitch     &      2 mm  \\
half gap            &      6 mm  \\
cathode strips    &     9-mm width with 1-mm spacing  \\
effective area     &  \\
~~C2        &    $16 \times 56$ cm$^2$  \\
~~C3         &   $20 \times 64$ cm$^2$  \\
~~C4         &   $20 \times 72$ cm$^2$  \\
gas mixture            &   Ar:ethane 50:50  \\
readout     &      x,y cathode strips  \\
\hline
\end{tabular}
\end{table}
\end{center}

Signals from each cathode strip were amplified in two stages.
The first was the preamplifier mounted on the MWPC which drove 
the cable going into the counting house where the signal 
was  further amplified by
the main amplifier. After the main amplifier, charge information
of each cathode was digitized in an ADC, after being multiplexed
 with
signals from six chambers into
one output in order to reduce the number of ADC channels.

The MWPC hit position was obtained using the
``charge-ratio method''\cite{chi1}. 
In this method, the analog signals of  three strips ---
the strip with the maximum charge and the two adjacent
strips --- are used.  Using data, the
value
\begin{equation}
\label{eq:char_mod}
R=\frac{q_{max}-q_{max+1}}{q_{max}-q_{max-1}}
\end{equation}
\noindent is calculated, where $q_{max}$, $q_{max+1}$, $q_{max-1}$ 
are the maximum induced
charges in the three strips, and tabulated.  The hit position is then
estimated by table lookup of values of $R$ determined
for 2-$\mu$m steps across the 10-mm wide strip pitch using
the induced charge distribution as in eq.(\ref{eq:char_mod}).

\begin{figure}
\epsfxsize=\linewidth
\begin{center}
\epsffile{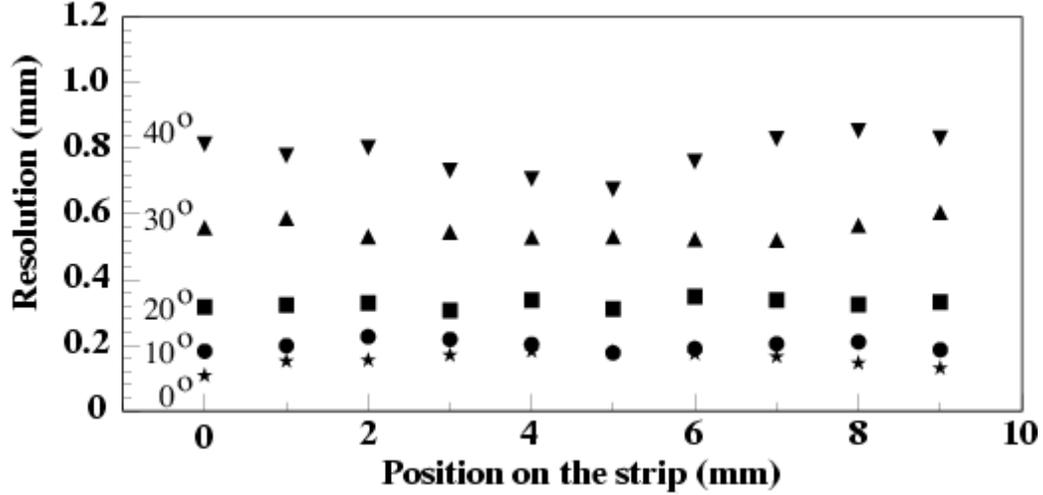}
\end{center}
\caption{MWPC position resolution ($\sigma$) as a function of position on
the cathode strip for different particle incident angles.}
\label{fg:chamresol}
\end{figure}

Using the charge-ratio method
there are two major problems which cause systematic shifts of
the reconstructed position.
One is caused by oblique incident particles.
In our case the incident angle could be as
large as 40$^{\circ}$. In such a steep incident angle, the gas
multiplication is no longer point-like. Rather, it is distributed on
the anode wire along a limited path. As a result, the induced
charge distribution differs from that in normal incident tracks.
To manage the oblique incident case, charge-ratio tables for various
incident angles were prepared and used in the analysis.
This treatment reduced
the systematic position shift  from $\pm$ 600  $\mu$m to
$\pm$ 50 $\mu$m.
 The position resolution ($\sigma$) was  measured as a function
of incident angle and is shown in Figure \ref{fg:chamresol}.

The other problem is caused by non-uniformity of the amplifier gains. 
From test results the maximum fluctuation of
the gains was as large as 20$\%$ and would result in a $\sim$
500 $\mu$m systematic shift in the reconstructed position.
To minimize this effect, gain calibration using ADC pedestal
information was  applied for each amplifier channel\cite{ty95}.
Using the calibration, the hill-valley structure seen in
 the reconstructed position  caused by non-uniform
 amplifier gain was removed and the overall position resolution was
improved about a factor of two.

\subsubsection{Gas recycler}

\begin{figure}
\epsfxsize=\linewidth
\begin{center}
\epsffile{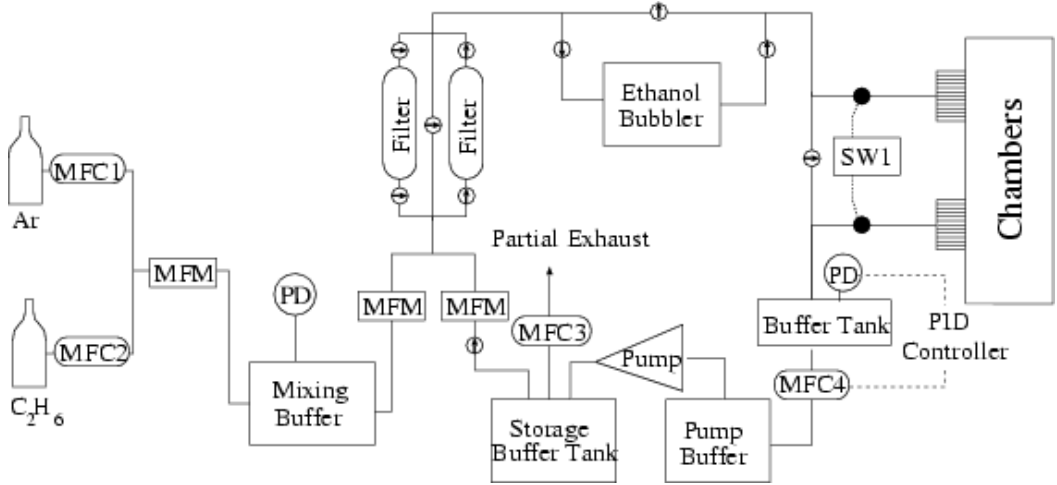}
\end{center}
\caption{Schematic diagram of the MWPC gas system.}
\label{fg:gassys}
\end{figure}

The recycler was designed to provide gas flows up to  3.5 
$\ell$-min$^{-1}$ with recycle ratios from 0\% to 100\%.  Reusable active 
filters in the system reduce water vapor and oxygen concentrations to below 
5 ppm. 
A schematic diagram of the system is shown in Figure \ref{fg:gassys}.

Mass flow controllers (MFCs) are set to deliver 
argon and ethane to prepare the required mixture at a high flow rate(~5 $\ell$-min$^{-1}$
Ar/ethane 50:50).  The flow of the mixture is monitored by a mass flow meter (MFM) 
and the mixed gas flows into a mixing buffer.  The pressure of the mixing 
buffer measured by a pressure transducer (PD) controls the MFCs, 
{\it i.e.}, the MFCs
are turned on whenever the pressure in the buffer falls below the lower set 
point and are turned off when the pressure rises above the upper set point. 

The gas supplied from the mixing buffer combines with the recirculated gas 
from the pump and flows into one of two filtration cartridges. Each cartridge 
contains one third 4-\AA~ molecular sieve to remove water, and two thirds 
BASF R3-11 activated copper for removal of oxygen. A particulate filter 
removes all particulates larger than 
2 $\mu$m. After filtration the gas flows through the 
temperature-controlled (typically -15 C) ethanol bubbler
which adds ethanol if there is too 
little in the gas stream and removes ethanol if there is too much.  

The gas 
then flows through a manual valve and a solenoid valve to the distribution rack.
To protect the chambers from excessive over or under pressure, the solenoid 
valve will automatically close if the pressure measured by 
the pressure transducer moves beyond the preset upper and lower set points.

The returned gas from the chamber distribution system flows to a return buffer tank.  
The pressure with respect to atmosphere at this point is controlled by 
a pressure transducer, PID controller, and mass flow controller and remains 
constant at the preset value of the PID controller. 

The distribution rack allowed the gas supplied from the recycler to be 
partitioned into 17 separate gas circuits, then recombined into a single 
return flow back to the recycler.  Manual flowmeters (rotameters) and mass 
flow controllers were used to control and measure supply flows in each circuit,
and mass flow meters measured return flows from the chambers.

Gas contamination was checked periodically with
a gas chromatograph\footnote{Shimazu Model GC-4C, Kyoto, Japan.}. 
When a filter
was saturated and no longer able to absorb oxygen, a regenerated one was 
put into service
by redirecting the gas, and the saturated one was removed for regeneration
in a controlled heating cycle in an 
H$_{2}$/N$_{2}$ gas flow.

\subsection{Time of flight counters}
\label{subsec:tofctrs}

Charged particles were identified by measuring time of flight ($TOF$)
between the fiducial counter next to the target (see Figure \ref{fg:tgtxsec}) 
and the TOF2 counters placed at the exit of the magnet gaps, just behind the 
C4 chambers, as shown in Figure \ref{fg:e246det}. 
Ideally for $TOF$ measurement, 
scintillation light should be collected from both ends of the scintillator
to cancel the
time jitter due to the light propagation time. However, the fiducial
counters 
are restricted to one-side readout because of limited space around the 
target. Therefore, we had to correct for the time walk related to the particle hit
position.
The 20-cm axial length of the fiducial counters was chosen
to match acceptance of the spectrometer and the stopping distribution  
of the 660-MeV/$c$ kaon beam with momentum spread of $\sim \pm 3 \%$.

The size of the TOF2 counter was 20 cm in width, 2 cm in thickness, and
80 cm in length. The scintillation light was read from both ends. 
Since there was a relatively strong
fringing field at the exit of the toroidal magnet, the phototubes had to be
located remotely ($\sim 50$ cm from the scintillators). The  scintillation
light was reflected once and transported to the phototube (Hamamatsu
H1161) through acrylic light guides. 
Output signals of  the fiducial counters and TOF2 counters were
split  by an analog divider
to feed the leading edge discriminator and  the charge
ADC. For timing adjustment, the discriminator output signal was
delayed and fed to 25-ps bin TDC.  

\begin{figure}
\epsfxsize=\linewidth
\begin{center}
\epsffile{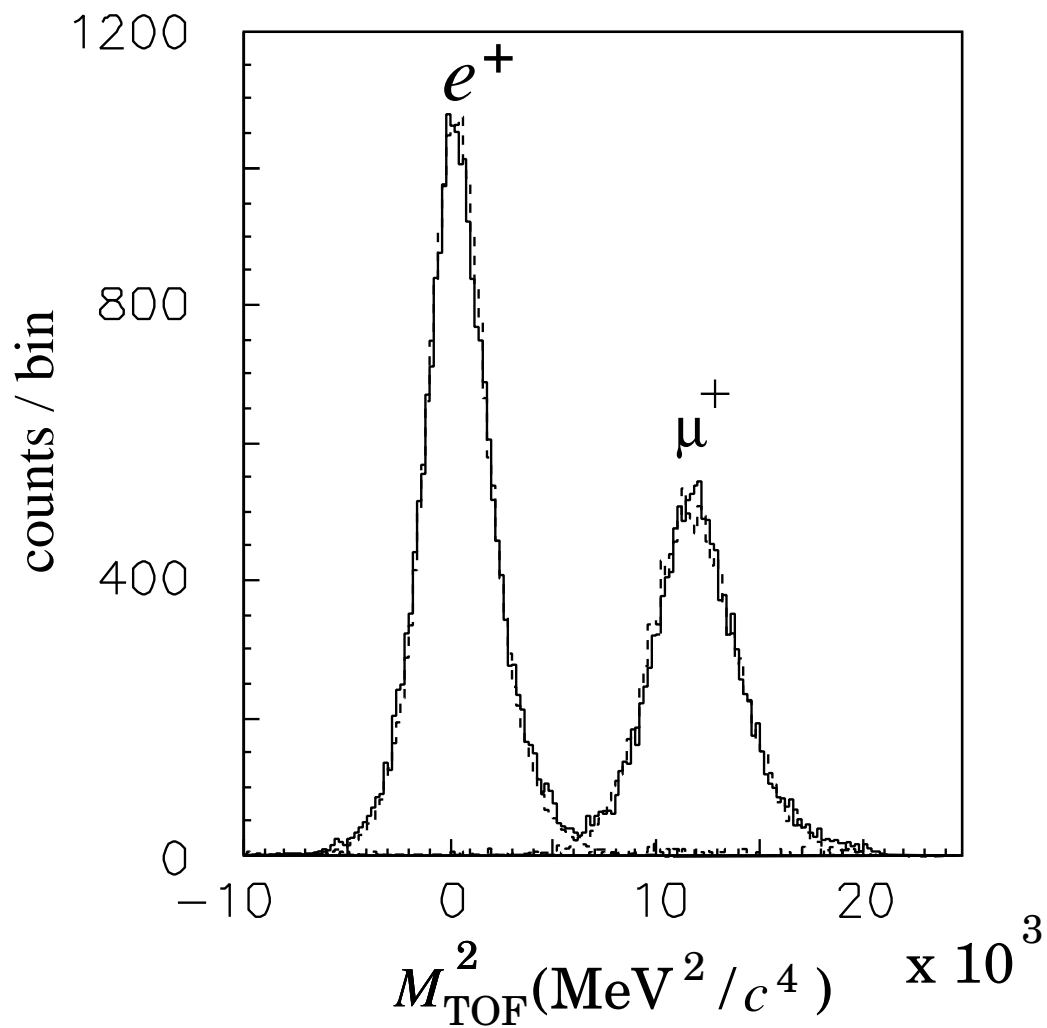}
\end{center}
\caption{Mass squared spectrum $M^{2}$ obtained in the TOF analysis.
 The solid line is 
the experimental data and the dotted line is a Monte Carlo simulation.}
\label{fg:massgmc}
\end{figure}

The hit time for the TOF2 counter ($t_{TOF2}$)  was
defined as the mean time measured by both
phototubes, while the time for the fiducial counter ($t_{f}$)
 was corrected to give
\begin{eqnarray}
 t_{TOF1}= t_{f}+x/c_n 
\end{eqnarray}
where $x$ is the particle hit position and $c_n$ is the speed of light in the 
plastic counter used to apply the photon propagation time correction.
The value of $c_n$=18 cm-ns$^{-1}$ was obtained in
a test experiment. A time-walk correction of the leading edge
discriminator was also applied. The actual $TOF$ between the fiducial and TOF2
counter was obtained using an offset term to account for fixed relative 
time differences between the counters. 
\begin{eqnarray}
 TOF = t_{TOF2}-t_{TOF1}+\Delta_t
\end{eqnarray}
The offset term, $\Delta_t$, was calibrated using the mono-energetic charged
particles of $\pi^+$ and $\mu^+$ from K$_{\pi 2}$ and K$_{\mu 2}$ decay. 
In these decays, the particle identification could be done
by analyzing their momenta, and therefore an actual $TOF$ was
calculated by using the information on flight path length and
momentum. Then, the velocity ratio of $\beta=v/c$ and mass squared of
the charged particle ($M^{2}$) were deduced as,
\begin{eqnarray}
        \beta=\frac{L}{TOF\times c}, \ \ M^{2}=p^{2}(1/\beta^{2} -1),
\end{eqnarray}
where $L$ is the flight length and $p$ is momentum of the charged
particle. Figure \ref{fg:massgmc} shows a typical $M^2$ spectrum integrated
for all momentum regions. The timing resolution is approximately 270 ps which 
is adequate to separate K$_{\mu 3}$ and
$K^+ \rightarrow \pi^0 e^+ \nu$  ($K_{e 3}$) decays.

\begin{center}
\begin{table}
 \caption{ Main parameters of the TOF measurement system.}
 \label{tb:tof}
 \begin{tabular}{lll}\hline
flight path  & \multicolumn{2}{c}{2 m (typical)}   \\
TOF resolution    &    \multicolumn{2}{c}{270 ps ($\sigma$) }   \\
   &   Fiducial-TOF1(start)  &  TOF2 (stop)  \\
\hline
scintillator type  &  Bicron BC408 & Bicron BC408 \\
PMT   &  Hamamatsu H3171-03  &  2 $\times$ Hamamatsu H1161  \\
dimensions  &  $27^*\times10^*\times200$ mm$^3$ & $200 \times 20 \times
800$ mm$^3$  \\
  &  $^*$ irregular (see Figure \ref{fg:tgtxsec}) &   \\
 \hline
 \end{tabular}
\end{table}
 \end{center}

\section{Photon Detector}
\label{sec:photondet}

The photon calorimeter was segmented into 768 CsI(Tl) crystals covering
a large
solid angle surrounding the kaon stopping target of approximately 70\% of $4\pi$ sr in order to measure the 
$\pi^{0} \to \gamma \gamma$ from $K_{\mu3}$ decay with high acceptance. 
The detector was optimized for the photon energy range of 10 -- 250 MeV and the high degree of segmentation provided good kinematic resolution for the $\pi^0$s. In order to reject background from other beam-associated 
accidentals, good timing resolution is necessary. The very limited 
space in the central part of the spectrometer magnet required a 
compact design with a read-out technique immune to the 
significant fringe magnetic field.
Details 
of the system have been described in ref. \cite{csinim} and 
\cite{csiele}. In this section the basic features are summarized.

\subsection{Barrel structure}
\label{subsec:barrel}

\begin{figure}
\epsfxsize=\linewidth
\begin{center}
\epsffile{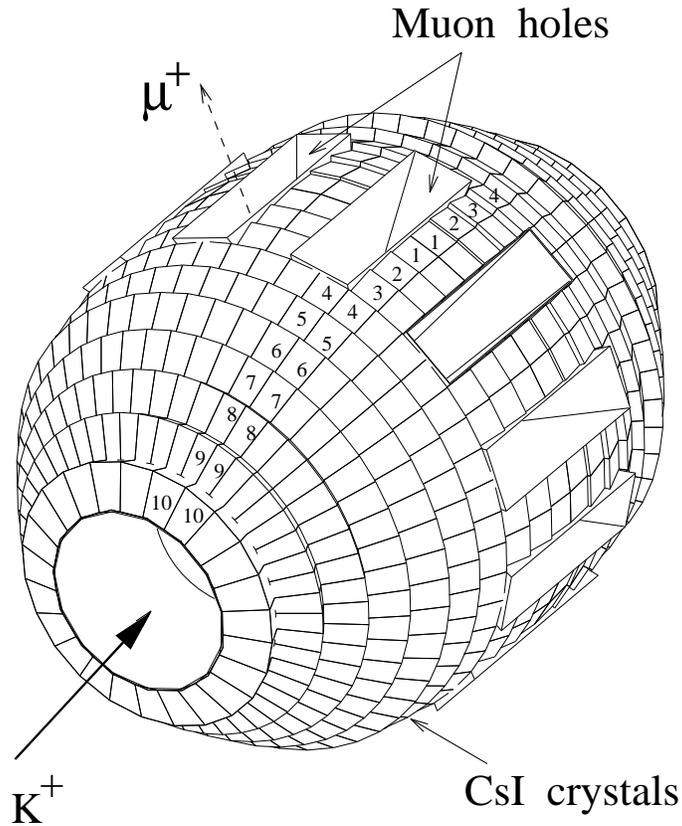}
\end{center}
\caption{Barrel structure of the CsI(Tl) detector. The numbers 1 -- 10 refer to the crystal types described in the text.}
\label{fg:csibarrel}
\end{figure}

Figure \ref{fg:csibarrel} shows the crystal arrangement in the detector. 
It has a barrel structure 
with two holes on the beam axis, and 12 azimuthal holes for muons to enter the 
spectrometer magnet gaps. The optimization of the crystal array to maximize overall acceptance was described in Section \ref{subsec:acceptopt}, and the diameter 
of the beam-axis holes was determined so as to accommodate beam 
instrumentation upstream, and the 
target and central tracking assembly 
downstream.  

The whole structure is upstream-downstream symmetric.
Around the target in the center of the barrel, the inner 
diameter is 40 cm and the outer diameter is 100 cm.  
There are 10 different crystal shapes (see Figure\ref{fg:csibarrel}) 
used as a function of
the polar angle position. The angular segmentation is 7.5 
degrees in both the polar and azimuthal directions except 
for the crystals nearest the 
beam axis (Type 10).
The axis of each crystal points to the center of the target
and has a length of 25 cm corresponding to 13.5 radiation lengths. 
The main parameters of 
the barrel are summarized in Table \ref{tb:csi}.

The crystal assembly is mounted in a cylindrical housing with high 
mechanical precision. 
An aluminum support structure 
bears the total weight of 1700 kg while minimizing material between the 
stopping target and the 
CsI(Tl) volume. Each crystal module is suspended from an 
outer rib structure avoiding 
a load on the inner 3-mm thick Al cylinder.  
Care was taken in the precise mounting to minimize gaps 
between crystal modules which would degrade energy resolution. 
The total assembly rolls on  4 small wheels on two rails
into position in the inner bore of the toroidal spectrometer magnet. 

\begin{center}
\begin{table}
\caption{Specifications of the CsI(Tl) photon detector.}
\label{tb:csi}
\begin{tabular}{ll}
\hline
Barrel & \\
~~segmentation             & $\Delta\theta=\Delta\phi=7.5^o$    \\
~~total crystal weight        & 1700 kg                         \\
~~inner diameter            & 40 cm                           \\
~~outer diameter            & 100 cm                          \\
~~detector length           &  141 cm                         \\
~~solid angle coverage       & 75\% of 4$\pi$                   \\
Modules & \\
~~crystal length               & 25 cm (13.5$X0$)              \\
~~surface treatment             & mirror polished(partially sanded)  \\
~~reflector                    &two layers of 120$\mu$ thick GSWP00010 \\
~~mean light yield (Type 1-9)        & 10920 p.e./MeV            \\
~~mean light yield (Type 10)         & 17400 p.e./MeV            \\
~~container                   & 100$\mu$ thick Al can          \\
~~PIN diode                   & $18 \times 18$mm, 300$\mu$ (Type 1-9)     \\
                                     & $ 25 \times 25$mm, 500$\mu$ (Type 10)    \\
~~power consumption           & 0.25W per preamplifier   \\ 
\hline
\end{tabular}
\end{table}
\end{center}

\subsection{CsI(Tl) modules}
\label{subsec:csimod}

\begin{figure}
\epsfxsize=\linewidth
\begin{center}
\epsffile{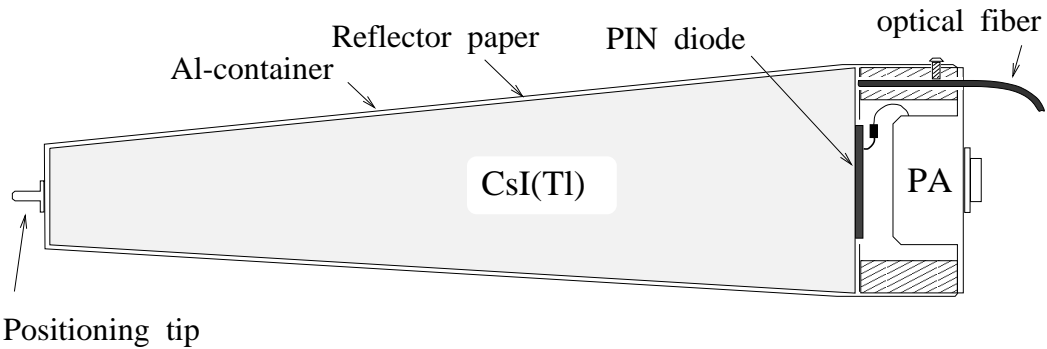}
\end{center}
\caption{CsI(Tl) crystal module assembly.}
\label{fg:module}
\end{figure}

The crystals were grown at the Institute of Single Crystals in Kharkov, 
Ukraine,
and shaped with a precision of 150 $\mu$m for each type. 
The structure of a crystal module is shown schematically in Figure \ref{fg:module}.
The typical size of the modules is $3\times3$ cm$^{2}$ at the front end 
and $6\times 6$ cm$^{2}$
at the back end. Each crystal was wrapped with two layers of 120-$\mu$m thick
white diffuse filter paper\footnote{Millipore Corporation, GSWP00010.}.
and contained in a 100-$\mu$m thick Al shell. To achieve good
light-yield uniformity along the crystal axis, before wrapping 
the mirror-polished surface of the crystals 
was ``tuned'' in the region nearest the readout by roughening 
by sanding. The light yield uniformity is as good as $\pm 2$ \% which 
is sufficient to provide linearity in energy  measurement. 

PIN photo-diodes of the type S3204-03 (18$\times$18 mm$^2$, 300-$\mu$m thick wafer) 
and S3584-05 (28$\times$28 mm$^2$, 500-$\mu$m thick wafer) were chosen for Type 1-9 
crystals and the larger Type 10 crystals, respectively
\footnote {Hamamatsu S3584--05 ( $C_{PD}$=200~pF,  $I_{D}<$100~nA) and S3204--03 ( $C_{PD}$=140~pF,  $I_{D}<$20~nA).}.
They were glued directly to the rear 
face of the crystal using a single component type SE777 silicone glue
\footnotemark[3] 
with short curing time and excellent adhesion between the crystal and diode. A 
charge sensitive preamplifier was mounted just behind the diode 
attached to an Al frame. 
Power consumption is 0.25 W for each preamplifier for a total detector
heat load of 200 W, which is cooled by temperature-controlled 
dry air flow into the barrel 
cylinder. The outside of each Al shell was painted black to obtain electrical 
isolation and absorb any stray reflected light.
An optical fiber feeds calibration light to the back end from a Xe-lamp 
monitoring system. 

Average light yields for the modules are summarized in Table \ref{tb:csi}
and almost all the modules have light yield more than 8000 photoelectrons
(p.e.)/MeV. 
The energy resolution of the modules was measured using
a 1.27-MeV $^{22}$Na $\gamma$ source to be 13-14\%(FWHM), and the average
equivalent noise level (ENL) was determined to be 63 keV. 

\subsection{Readout electronics}
\label{subsec:csielec}

Each low-noise pre-amplifier feeds a main amplifier (MA) 
with low-gain and high-gain outputs and incorporated Timing-Filter-Amplifier
(TFA).
 The low-gain signal is fed to a
peak-sensing ADC as well 
as to a transient digitizer (TD) for energy measurement. 
The TFA provides a 450-ns wide pulse 
to a 
constant-fraction discriminator (CFD) whose output signal is fed to a 0.7-ns bin TDC for time measurement. 
The gain and time constant of the preamplifiers 
were optimized for the average energy deposit and counting rate. The adjustable
shaping time constant of the MA was typically 1.0$\mu$s, and a 
pole-zero cancelation circuit and baseline restorer in the main amplifier 
allowed good high rate 
performance; at 10 kHz average rate, energy resolution did not degrade 
appreciably, keeping 92\% counting efficiency.

The TD is based on a switched-capacitor-array (SCA)
integrated circuit which allows waveform 
recording with 12-bit accuracy at the maximum sampling rate of 10 MHz\cite{sca}.
It provided additional double-pulse resolving ability 
at higher rate, supplementing the ADC and TDC 
systems. The main characteristics are summarized in Table \ref{tb:td}.

\subsection{Xenon lamp monitoring system}
\label{subsec:xelamp}

The Xe-lamp monitor system provided test light pulses 
to each module to detect failed photo-diodes 
and to do other diagnosis of the  readout electronics.
The light is distributed to each crystal module
through an optical glass fiber and a 
mixer fanout box. The light signal is split in two stages, 
24 $\times$ 32 to feed the 768 crystals.
The 
distributors provide uniform light to each crystal to within $\pm12\%$ 
with overall loss of 18dB. The Xe lamp \footnote{Hamamatsu L2189}
 operates at a repetition rate of 
10 Hz, and has an emission 
spectrum similar to CsI(Tl). 
The typical response of a crystal corresponds to an energy 
deposit of 50 to 100 MeV. The short term stability of pulse height was 2\% (FWHM) 
which was monitored by a PMT which also provided the trigger signal in the 
calibration measurement.   

\begin{table}
\caption{Main parameters of the transient digitizer.}
\label{tb:td}
\begin{tabular}{ll}
\hline
type                        & switching capacitor array + ADC \\
bin size                       & 640 ns                     \\
number of bins                 & 128                        \\
dynamic range                  & 12 bit                     \\
conversion time              & 820$\mu$s for 4 multiplexed SCAs \\
linearity                     & $<$0.1\%                      \\
\hline
\end{tabular}
\end{table}

\subsection{$\pi^{0}$ detection performance}
\label{subsec:pi0perf}

The CsI(Tl) calorimeter performance is summarized in Table \ref{tb:csires}.
The energy resolution was measured 
using  tagged photon beams with variable energy in a prototype assembly of 
30 similar crystals
assembled in a $5\times 6$ matrix.
Energy resolutions of $\sigma_E/E=$ 2.8\% and 4.3\% were obtained for 200 MeV and 
100 MeV photons, respectively. In the actual detector with the 12 muon 
holes, however, the
resolution is dominated by lateral leakage of a shower resulting in a
 low-energy tail. 
 Typical resolution for the sum 
energy  of two photons from $K_{\pi2}$ decay is $\sigma_E/E=$4.1\%. 
The single crystal resolution of $\sigma_E/E=$1.7\% 
was measured using the 152-MeV $\mu^+$ from $K_{\mu2}$ decay
 by selecting events with no energy 
deposit in the neighboring crystals. 
The $K_{\mu2}$ muons were used for the crystal gain calibration. The 
electronic 
noise contribution to the energy resolution was found to be negligibly small;
from an ADC 
pedestal measurement the incoherent contribution was $\sigma_E=$250 keV 
when summed over 9 crystals while the coherent contribution 
was $\sigma_E=$11 keV per module.

\begin{figure}
\epsfxsize=\linewidth
\begin{center}
\epsffile{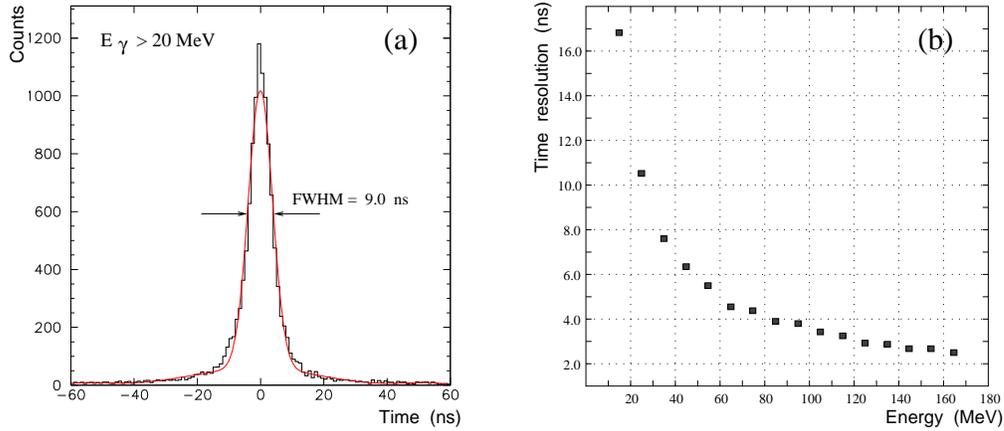}
\end{center}
\caption{Timing characteristics of the photon detector. (a) Time spectrum of all CsI crystals 
(center of clusters) with aligned peak position for $K_{\mu3}$ events. (b) Time resolution 
($\sigma$) as a function of photon energy.}
\label{fg:time}
\end{figure}

For $\pi^0$ kinematics determination, the photon spatial resolution is essential. 
The directional
resolution of $\pi^0$s was $\sigma_\theta$=2.3 degree using a back-to-back
correlation between the $\pi^0$ and $\pi^+$ of $K^+\rightarrow\pi^+\pi^0 (K_{\pi2})$ decay, 
corresponding to a spatial resolution of 7.6 mm. A
typical time spectrum is shown in Figure \ref{fg:time}(a). For the energy range of 10-200 MeV
the resolution was $\sigma_T=$ 3.8 ns in the experiment. Figure \ref{fg:time}(b) 
shows the energy dependence of the resolution. 

\begin{figure}
\epsfxsize=\linewidth
\begin{center}
\epsffile{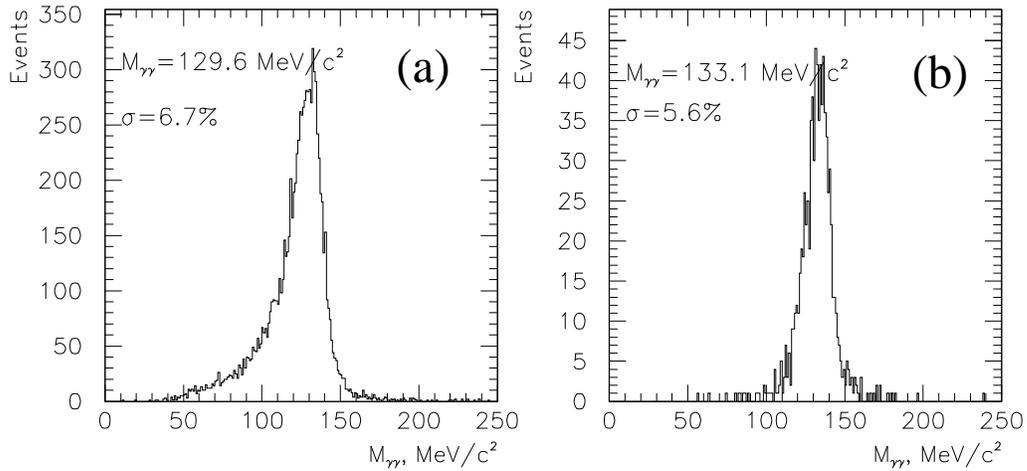}
\end{center}
\caption{Invariant mass spectra of two photons from $K_{\pi2}$ using $5\times5$ clustering:
(a) all events reconstructed in the detector; (b) events detected away from the muon hole area.}
\label{fg:mgg}
\end{figure}

Figure \ref{fg:mgg} shows the invariant 
mass spectrum of two photons from $K_{\pi2}$ decay using $5\times 5$ 
crystal clustering:
(a) is the spectrum of the whole detector with $\sigma_{M_{\gamma\gamma}}$= 6.7\% at 129.6 
MeV/$c^2$ which includes the
deterioration due to shower leakage. The resolution improves to 5.6\% for events 
away from the muon holes (b) \cite{marat}.

\begin{table}
\caption{Performance of the CsI(Tl) photon detector.}
\label{tb:csires}
\begin{tabular}{ll}
\hline
energy resolution $\sigma$/E &          4.3\% at 100 MeV      \\
                        &          2.8\% at 200 MeV      \\
electric noise (incoherent)   & 250 keV ($\sigma$) for 9 modules    \\
electric noise (coherent)     &          11keV per module       \\ 
linearity                 &       2\%  for  5 -- 250 MeV     \\
spatial resolution (rms)      &          7.6 mm at 200 MeV     \\
angular resolution (rms)     &         2.2 -- 2.4 degree          \\
time resolution (rms)        &          3.5 ns at 100 MeV      \\
$\pi^0$ mass resolution      &             $\leq 7.5$ MeV/$c^2$      \\
\hline
\end{tabular}
\end{table}

\section{Muon Polarimeter}
\label{sec:polarim}

By using the longitudinal field method introduced in Section \ref{subsec:stoppedk},
the transverse polarization component $P_{T}$ is 
preserved under the holding field and the in-plane components $P_{L}$ 
and $P_{N}$ are
precessed as illustrated in Figure \ref{fg:msr}. 
The transverse component is then detected as a left-right,
or clockwise-counter-clockwise
asymmetry of decay positrons with respect to the muon stopper. 

\begin{figure}
\epsfxsize=\linewidth
\begin{center}
\epsffile{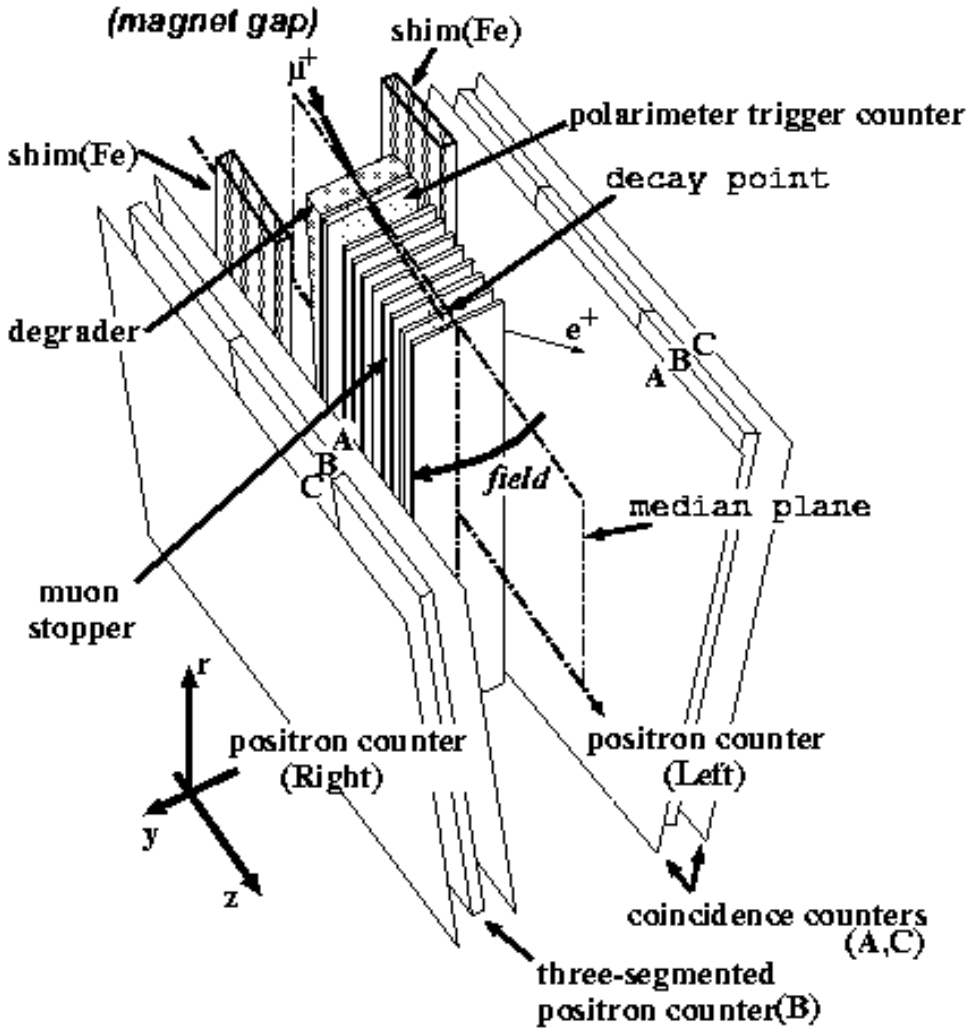}
\end{center}
\caption{Schematic structure of the muon polarimeter.}
\label{fg:polarimeter}
\end{figure}

The structure of the polarimeter is shown in Figure \ref{fg:polarimeter}. The field 
on the stopper was formed by iron shim plates which shaped the 
superconducting magnet fringing field. Its distribution on the muon 
stopper was symmetric but not uniform as can be seen in 
Figure \ref{fg:field}.

To detect a tiny effect of $P_T$ 
the analyzing power $A_T/P_T$ should be as large as possible, and 
there must be no depolarization of the  stopped muon spin. 
Any depolarization which could be material and field strength dependent not
only reduces the effect but also has the potential to introduce
spurious effects.

\subsection{Rotation of the polarization vector in flight}
\label{subsec:polrot}

As the muon goes through the magnetic field in the magnet gap 
its spin rotates around the B field axis along with the
trajectory. The effect of the spectrometer magnetic field on the rotation of the muon polarization vector  was studied in a GEANT3 simulation \cite{geant} which included the relativistic Bargmann-Michel-Telegdi equation \cite{BMT} to treat polarization dynamics. Muon depolarization due to multiple scattering was evaluated \cite{lyuboshits} separately and found to be unimportant.
After being bent by about 90 degrees through the
gap, the projection of polarization  on the median plane of the gap is
conserved as well as the median-plane perpendicular component.
By integrating forward or backward pion regions the net
perpendicular component, which should be $P_T$, remains the
azimuthal polarization at the exit of the gap. 

On entering the
Cu degrader and Al stopper,  depolarization takes place due to
the relativistic effect in multiple scattering. The degree of
depolarization
can be evaluated both for longitudinal component ($P_L$) and
transverse components ($P_N$ and $P_T$)\cite{lyuboshits} and was
found to be negligibly small.  The spin dependence of Mott scattering was
also considered, since it may generate
left-right asymmetry in the muon stopping distribution giving
an opposite effect for forward and backward pions potentially
resulting in a spurious T-violating asymmetry.  However, this
asymmetry is very small for small scattering angles in
multiple scattering where the cross section is large.

\subsection{Spin rotation field}
\label{subsec:spinrot}

\begin{figure}
\epsfxsize=\linewidth
\begin{center}
\epsffile{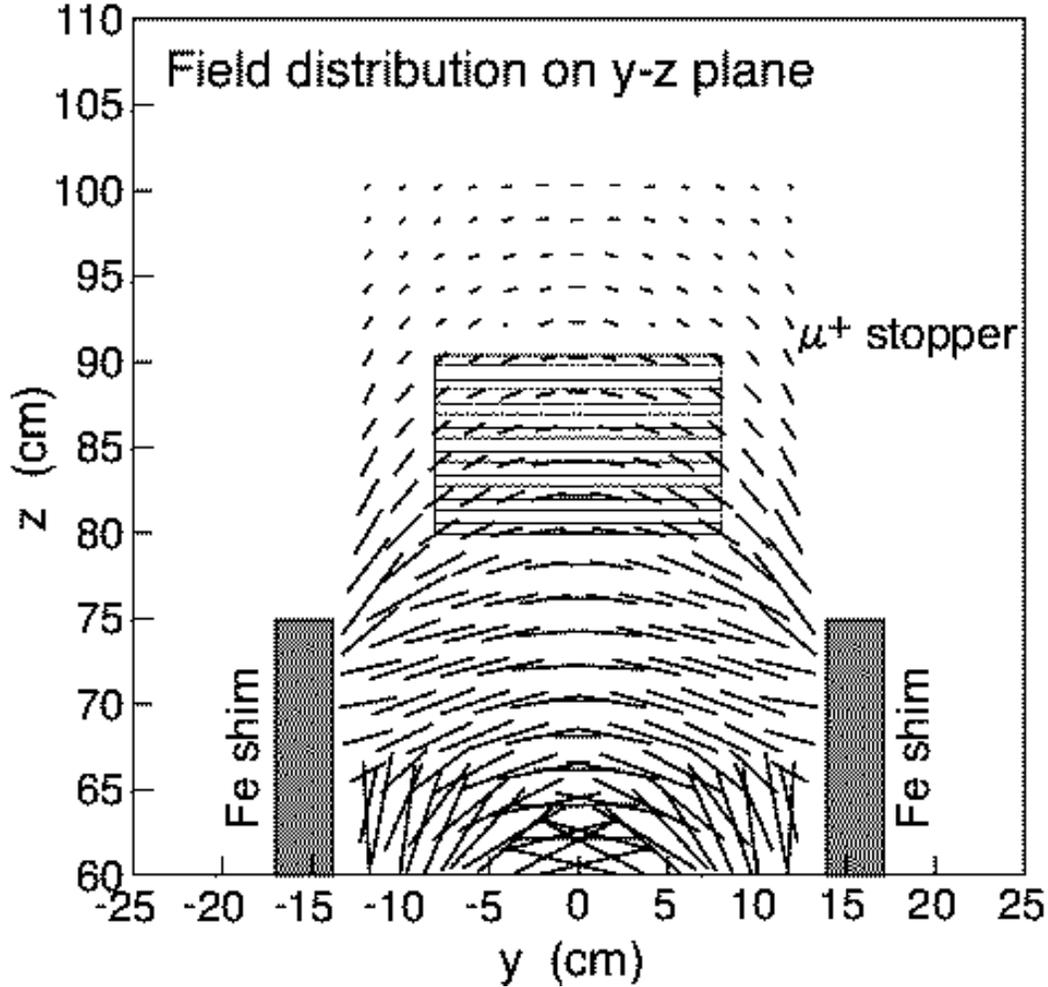}
\end{center}
\caption{Magnetic field distribution on the muon stopper.}
\label{fg:field}
\end{figure}

The configuration of the muon degrader, stopper, positron counters, 
and magnetic field shim plates 
was designed to
optimize the 
polarization sensitivity and minimize spurious instrumental effects. This
required a careful balance between ensuring the necessary
solid angle for positron detection and avoiding asymmetries in 
 the field distribution. 
The field distribution the in $r-\phi$ 
plane is shown in Figure \ref{fg:field}. If seen locally $P_N$ and 
$P_L$ are admixed to the azimuthal component $P_T$  in the course of precession 
around the tilted field line. It was therefore essential to realize a 
perfect symmetry of the field distribution across the magnet-gap median
plane so that any effect from the in-plane components vanishes 
in total.

The position and structure of the
shim plates was designed using a three dimensional field calculation
code TOSCA\cite{tosca}. 
The field on the muon stopping distribution should be as high as
possible by guiding the fringing field from the superconducting 
magnet so that it decouples the stray fields in the
experimental area. In addition a possible effect from coil misalignment 
in the cryostat should be negligible.
There should be no saturation in the iron in order to suppress
unwanted effects which might be material dependent and have hysteresis
characteristics. Finally, a simple structure was prefered for ease of precise
installation.

In the designed configuration an average field strength in the region 
of stopping muons
of 130 Gauss was realized for the 0.9-T magnet excitation. The
magnetization in the shim plates made of high quality 
magnet steel is 300 Gauss maximum and well below saturation.   

Each plate was manufactured with a precision of 50 $\mu$m and the 
installation was done with an accuracy of 200 $\mu$m relative to the 
magnet gap median planes by using a specially prepared jig. Field 
mapping was performed at 0.9-T excitation  using a 3-dimensional 
Hall element mounted on a specially made scanning device. The 
four-Euler-angle method was employed. Details of the measurement in which 
the strength accuracy of 0.1 Gauss and an angular accuracy of 1.0 mr 
are achieved was described in  \cite{fieldmap}. 
Three tilt angles of the distribution were deduced from the measured field maps,
 and 
are summarized in Table \ref{tb:fieldtilt}. These 
characteristics of the muon spin rotation field are adequate for 
the present experiment.

\begin{table}
 \caption{Magnetic field characteristics on the muon stopper.}
 \label{tb:fieldtilt} 
 \begin{tabular}{ll}
   \hline
   average field strength       & 130 Gauss                 \\
   accuracy of field mapping    &                           \\
   ~~ field strength            & 0.07 Gauss                \\
   ~~ field vector angle        & 1.0 mr                    \\
   asymmetry of the distribution&                           \\
   ~~ offset in $y$ direction   & 0.9 mm                    \\
   ~~ rotation around $z$ axis  & 0.15 mr                   \\
   ~~ rotation around $x$ axis  & 1.58 mr                   \\   
   \hline
 \end{tabular}
\end{table}

\subsection{Al stopper and material test}
\label{subsec:alstop}

The muon stopper was required to be thick enough to stop muons, but it must 
also allow
the positrons to escape.
An array of eight 6-mm thick pure aluminum slabs, 
160-mm wide by 550-mm long, with 8-mm
spacing was used. 
The stopping efficiency for $K_{\mu 3}~$ muons was 78\%.
The material we used was type 1N99 which corresponds to a purity of
more than 99.99\%. 
It is well known that muons imbedded in pure Al show neither
intitial loss of polarization nor spin relaxation\cite{aldepol}.
Actual material was 
studied in  a $\mu$SR experiment with polarized muons at the
Meson Science Laboratory of the University of Tokyo (UT-MSL).
Muons from the beam were stopped and
allowed to decay in the aluminum located in the center of 
two Helmholtz coil magnets, allowing the application of both
transverse and longitudinal magnetic fields of variable strength.
Located symmetrically upstream and downstream from the stopper were 
two arrays of positron detectors.

\begin{figure}
\epsfxsize=\linewidth
\begin{center}
\epsffile{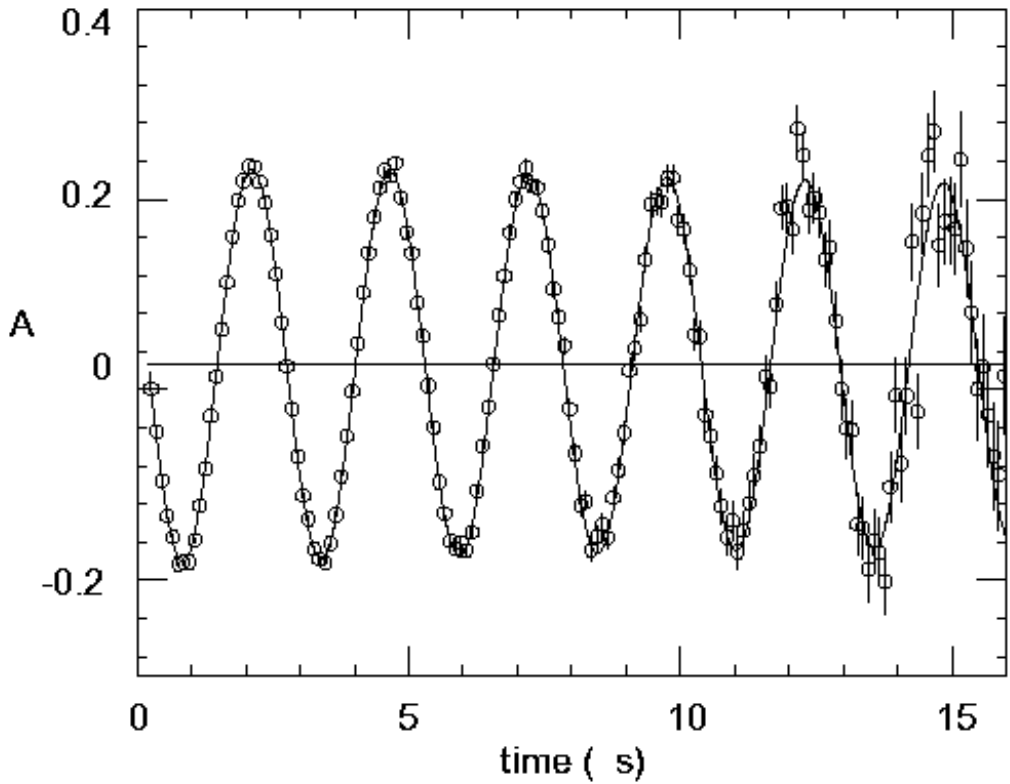}
\end{center}
\caption{The $\mu$SR spectrum in a 30-G transverse field from a sample of type 1N99 pure aluminum used for the polarimeter muon stoppers. The vertical axis is the asymmetry after subtraction of the muon decay lifetime.}
\label{fg:alummusr}
\end{figure}

The forward/backward asymmetry of muon decay positrons
was determined as the ratio
\begin{equation}
\label{eq:musrratio}
A =\frac{ N_{b} - N_{f} } { N_{b} + N_{f} },
\end{equation}
 where $N_b$ and $N_f$ are the time-dependent positron counts
in the forward and backward counters, respectively. After
 subtracting constant background components in $N_b$ and $N_f$
and normalization, the asymmetry $A$ was fit to a function
\begin{equation}
\label{eq:musrfit}
F(t) = A_0 e^{-\lambda t}   \cos{(\omega t +\phi_0)},
\end{equation}
      where $A_0$ is the initial asymmetry, $\lambda$ is the damping
      rate determined by spin relaxation and field inhomogeneity,
      $\omega$ is the precession angular velocity, and $\phi_0$ is the
      small initial phase shift. Figure \ref{fg:alummusr} shows a typical
      asymmetry spectrum for a 30-Gauss field. The measured damping
      rates for three different field strengths are listed in Table \ref{tb:6:musrres}. 
They are all small enough to indicate that spin
      relaxation in this Al material is negligible.

\begin{center}
\begin{table}
 \caption{ Measurement of $\mu$SR amplitude damping rate for Type 1N99 Al 
($>99.99$\% pure) in various transverse field strengths.}
 \label{tb:6:musrres}
\begin{tabular}{ccc}\hline
Field (Gauss)   &   Initial asymmetry   &  Damping rate ($\mu{\rm s}^{-1}$) \\ \hline
30   &   0.208 $\pm$ 0.001   &   0.0062 $\pm$ 0.0012  \\
50   &  0.205 $\pm$ 0.001    &   0.0081 $\pm$ 0.0010  \\
70   &  0.199 $\pm$ 0.001   &    0.0051  $\pm$ 0.0010  \\ \hline
  \end{tabular}
\end{table}
 \end{center}

\subsection{Positron counter system}
\label{subsec:posctr}

\subsubsection{Structure}
\label{subsubsec:posctrstruct}

Two adjacent muon stopper arrays share one positron counter assembly
between them, being the clockwise counter with respect to one gap 
and the counter-clockwise counter of the neighboring gap.
Each of the 12  positron counter assemblies comprises a three-fold segmented positron
counter sandwiched between two coincidence counters, depicted as (B) and (A,C), 
respectively, in
Figure \ref{fg:polarimeter}. 
Each segment of the positron counter is 10-mm thick, 100-mm
wide, and 840-mm long, so that the total width is 300 mm. Each segment is read
by two mesh-type PMTs (Hamamatsu H6154), which retain high gain in a
magnetic field. 
The 3 $\times$
380 $\times$ 920-mm$^3$ coincidence counters were designed so that
their solid angles from any point in the muon stopper are larger than
that of the positron counter.
The triple coincidence among all three layers  confirmed a positron
free from the background, mainly photons and neutrons.
The geometrical acceptance of the positron counter assembly is about
13\% of 4$\pi$ solid angle.

Each element in the polarimeter was accurately positioned
to  better than 1 mm
in order to minimize false asymmetry in
the positron counting due to instrumental mis-alignment.
The maximum alignment deviations of the positron counters and 
muon stoppers are listed in
Table \ref{tb:poalignment},
 and are smaller than the calculated
limitations. 

\begin{center}
\begin{table}
 \caption{ Maximum alignment deviation of polarimeter elements.}
 \label{tb:poalignment}
  \begin{tabular}{cccc} \hline
& $\Delta x$ [mm]& $\Delta y$ [mm]& $\Delta z$ [mm]\\ \hline 
muon stopper & $\pm$1~~~& $\pm$0.2 & $\pm$0.2 \\ 
positron counter & $\pm$0.5& $\pm$0.2 & $\pm$0.2 \\ \hline 
  \end{tabular}
\end{table}
 \end{center}

\subsubsection{Coincidence}
\label{subsubsec:posctrcoinc}

\begin{figure}
\epsfxsize=\linewidth
\begin{center}
\epsffile{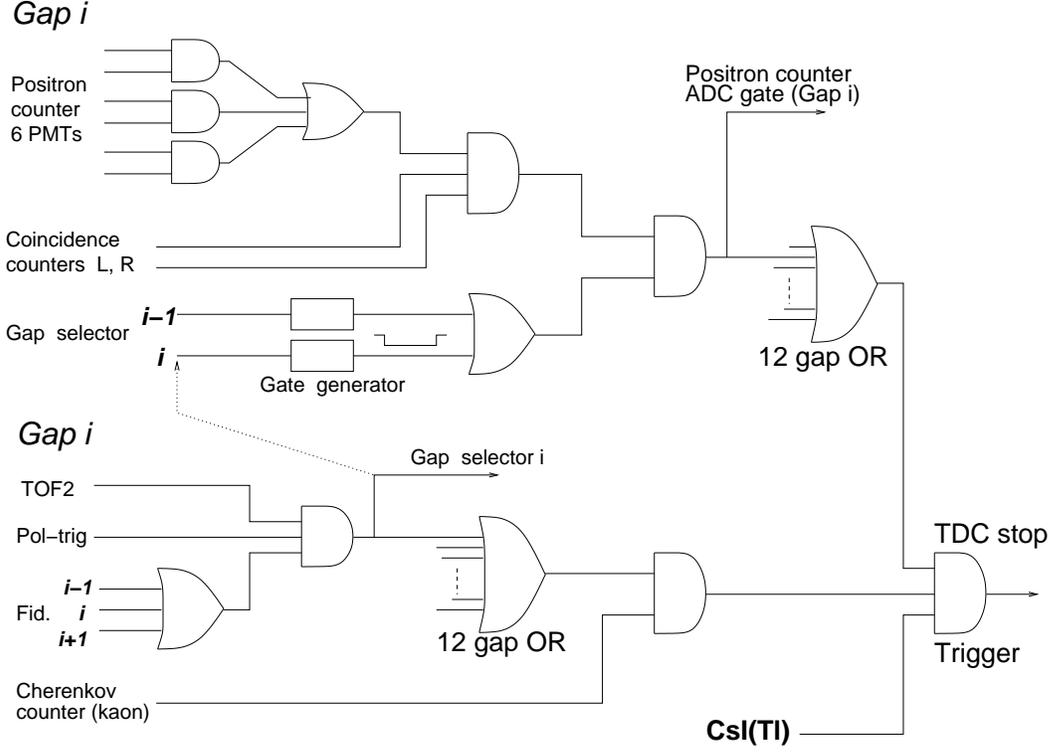}
\end{center}
\caption{ Block diagram of the positron detection logic.}
\label{fg:diagpos}
\end{figure}

A block diagram of the positron detection logic is shown in
Figure \ref{fg:diagpos}. 
In order to obtain the time spectrum of the positron from $K_{\mu 3}~$ 
muons, 
the coincidence information between the three-fold segmented positron
counter and the coincidence counters was used as the TDC stop signal.
The TDC start was generated by the coincidence of TOF2, the
polarimeter trigger counter and the fiducial counters.
To reject other decay modes, additional signals from other detectors
were available.
The discriminator thresholds of the counters were set to be sufficiently
low ($<$200 mV) in order to achieve close to 100\% efficiency for positrons. 
The triple coincidence reduced the count rate to about 40\% of that 
without the coincidence counters  by suppressing the
background of mainly neutral particles.
The PMT gain drifts were monitored 
using the  MIP peak position in the
ADC spectra.

\begin{figure}
\epsfxsize=\linewidth
\begin{center}
\epsffile{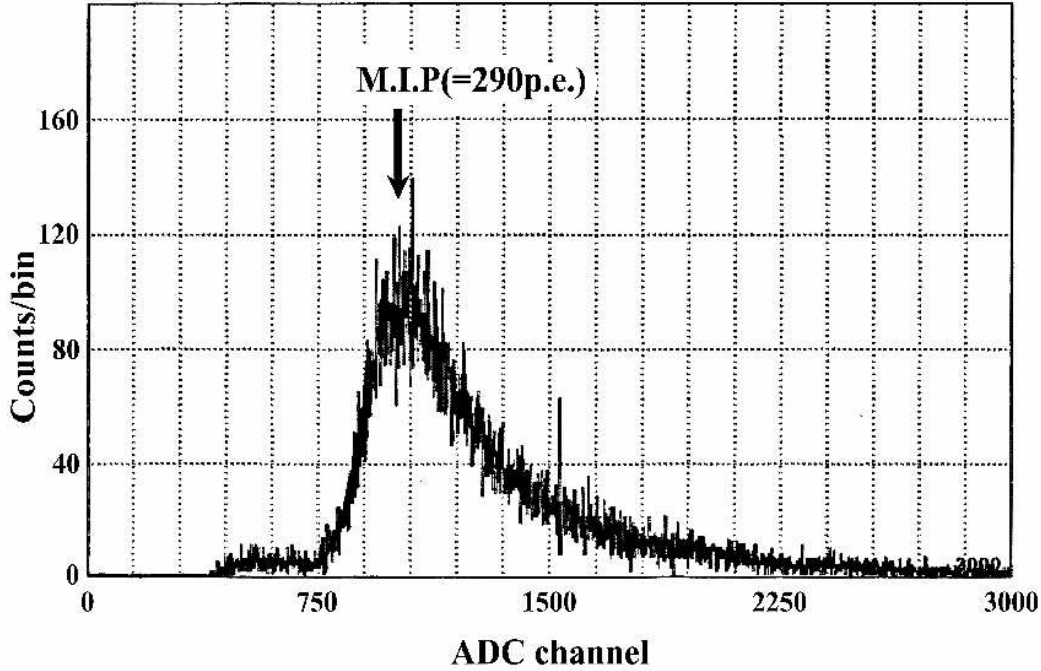}
\end{center}
\caption{ Positron counter ADC spectrum from cosmic rays.}
\label{fg:mip}
\end{figure}

Figure \ref{fg:mip} shows a typical ADC spectrum of 
the energy deposition of minimum ionizing cosmic rays in one positron counter,
corresponding to a light yield at the peak of 290 photoelectrons.
An additional calibration system using a Xenon lamp was also
employed. Optical fibers distributed the light to all 36 scintillators of 
the three-segmented positron counters. The system supplied a stable photon yield
to each PMT larger than that for MIP, with a fluctuation of  less than
10 \%.

\subsection{Asymmetry measurement and polarization}
\label{subsec:asympolmeas}

The measurement of muon polarization $P_{\mu}$ is based on the
asymmetry
of positrons emitted in muon decays with a lifetime of $\tau_\mu=2.2$ s.
 In
the longitudinal field method, the angular distribution of the energy
integrated positron counting rate, $N_{e^+}(\theta,t)$ is expressed as

\begin{equation}
          N_{e^+}(\theta,t)= N_0 \exp(-t/\tau_\mu) [1 + \alpha
P_{\mu}\cos{\theta}].
\end{equation}

Here, $\alpha$ is the analyzing power, $P_\mu$ is the longitudinal
muon polarization, and $\theta$ is the emission angle relative to
the axis. (For an ideal case of no interaction with matter of the emitted
positron and no trajectory distortion, $\alpha$ is 1/3.)
In the present configuration, the longitudinal component is the
average transverse polarization in the azimuthal direction, which
can be determined from clockwise/counter-clockwise ($cw/ccw$) asymmetry
of time integrated positron counting rates as,

\begin{equation}
      P_T =( \alpha \langle\cos{\theta_T}\rangle)^{-1} \frac{N_{cw} - N_{ccw}}{N_{cw} + N_{ccw}}.
\label{eq:alphapt}
\end{equation}

 The analyzing power, $\alpha$, included three experimental 
effects from (1) the finite solid angle of the
counters, (2) scattering and absorption in the Al stopper, and (3)
spin precession around the magnetic field which is not everywhere
parallel to the azimuthal direction (see Figure \ref{fg:field}). 
The analyzing power $\alpha$
could be optimized to provide the maximum figure of merit,
$\alpha \sqrt{N_{e+}}$. Experimentally, it was checked using the $K_{\mu3}$
data. By
selecting  $\pi^0$s
normal to the beam on the left side and right side, one orients the
in-plane $\mu^+$ polarization $P_N$ azimuthally in the  stopper
rather than $P_T$ in the case of $fwd$ or $bwd$ $\pi^0$s. 
The positron asymmetry $A_T$ was measured as the $cw$
and $ccw$
difference of the counting rate as in Eq. \ref{eq:alphapt}. The average $P_N$ was
calculated from a Monte Carlo simulation. $\alpha$ is then deduced
as
$\alpha = A_T / <P_T>$ to be typically 0.28 -- 0.30.

\subsection{Veto counter system}
\label{subsec:vetoctr}

\begin{figure}
\epsfxsize=\linewidth
\begin{center}
\epsffile{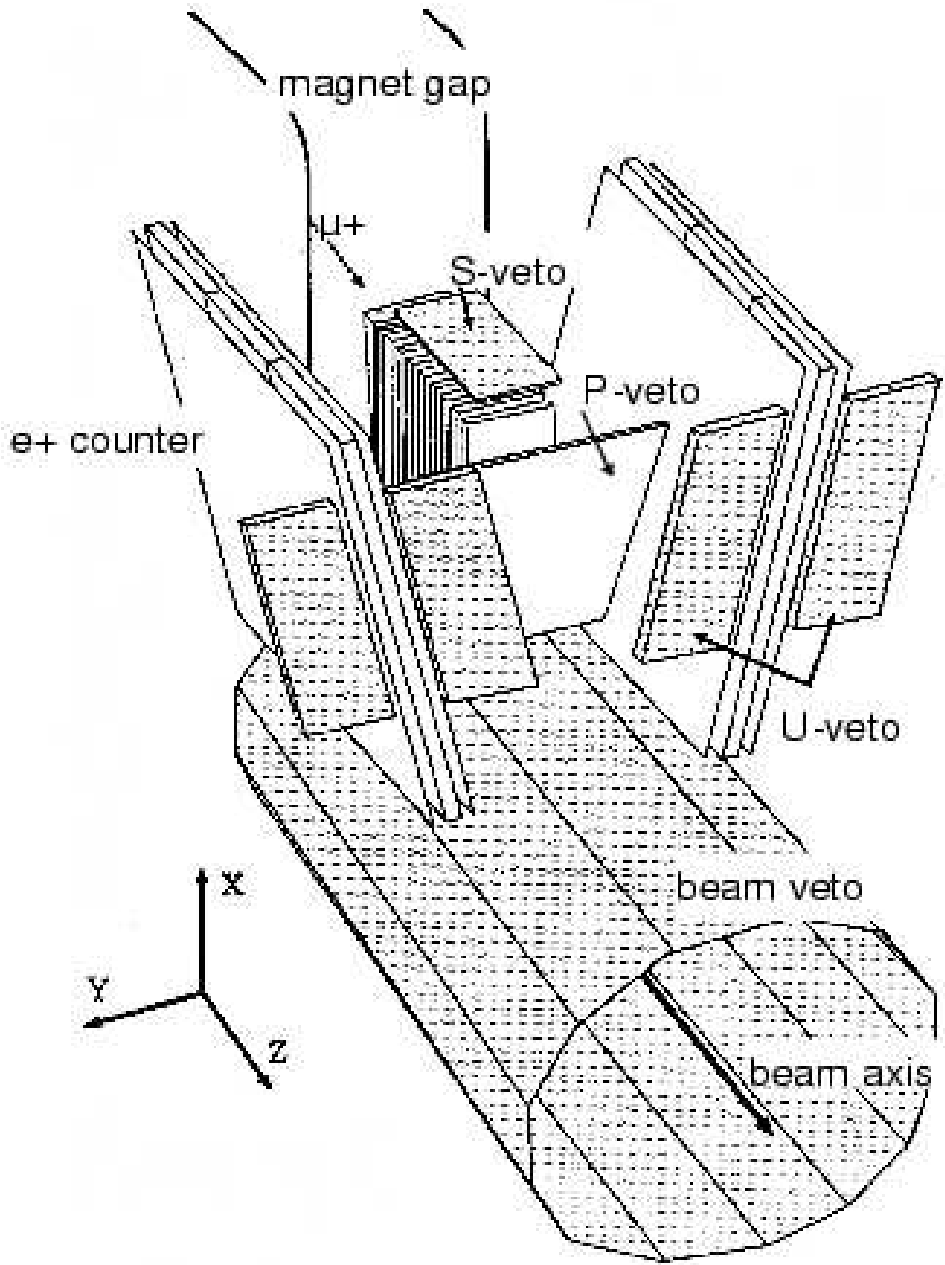}
\end{center}
\caption{The array of veto counters in the polarimeter.}
\label{fg:vetos}
\end{figure}

One of the advantages of the present experiment is 
a relatively low
background level at the polarimeter compared with the previous
in-flight experiments. However,  the influence of the backgrounds due to beam
particles cannot be eliminated completely. In order to reduce these
backgrounds and study their characteristics, an array of veto counters
was installed, as shown in
Figure \ref{fg:vetos}. 

The main components are the B-veto counters, which cover the 
inner-radius region of the
polarimeter to isolate the polarimeter from the
beam. The P-veto and U-veto counters counters located behind the muon stopper and
positron
counter, respectively, 
were aimed at rejecting events such as muons passing through 
or scattering off the
muon stopper.  The S-veto counters, covering the outer-radius region of
the
polarimeter, rejected muons exiting the muon stopper to
the outside of the polarimeter. About 30\% of the accidental
background in the positron time spectrum could be rejected by
these veto counters. Also, by using calibration data which were
collected with a special trigger,
we found that
10\% of $K_{\mu 3}$ decay muons escaped the polarimeter
and were rejected.

\section{Trigger}
\label{sec:trig}

\begin{figure}
\epsfxsize=\linewidth
\begin{center}
\epsffile{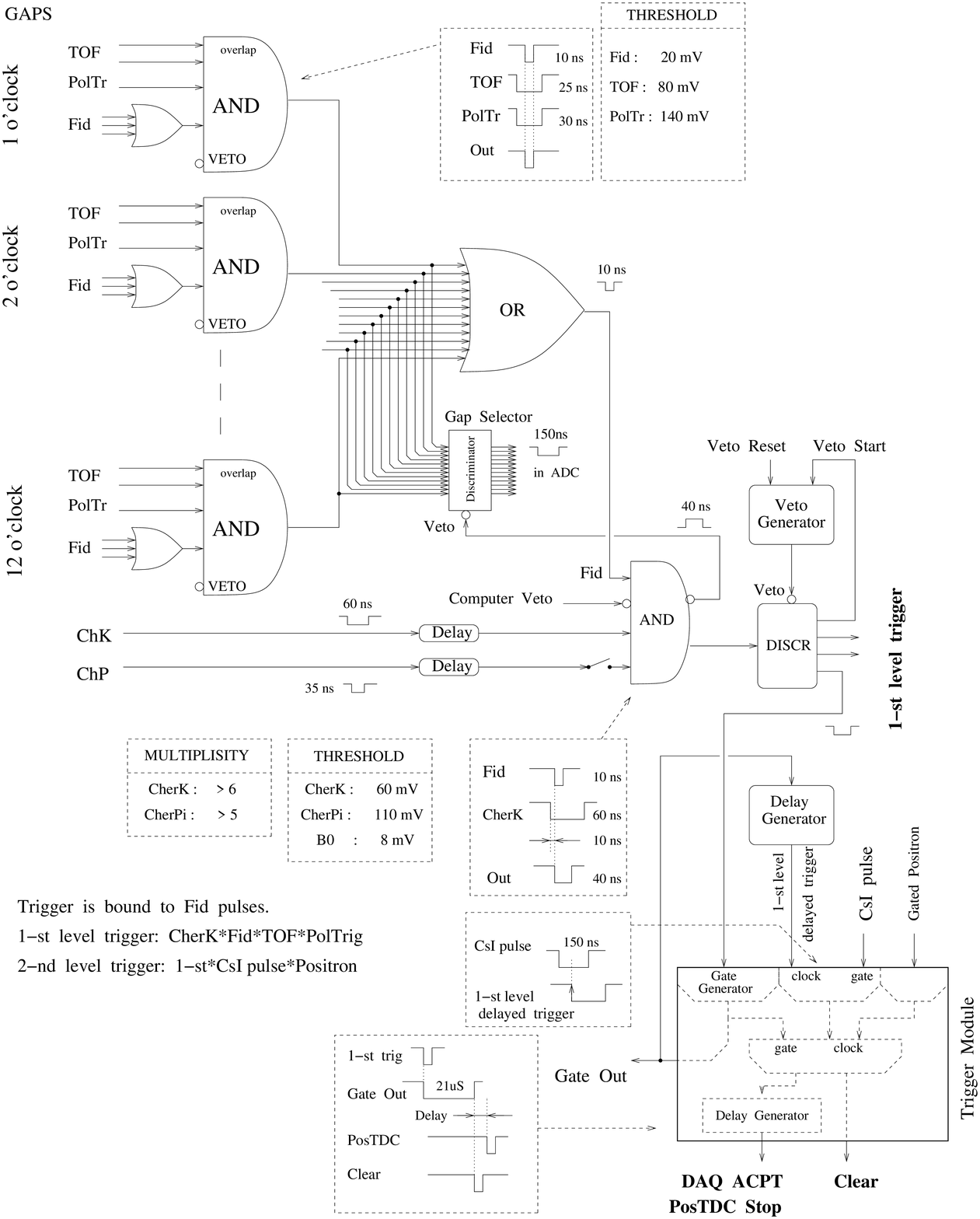}
\end{center}
\caption{Block diagram of the trigger logic.}
\label{fg:trigger}
\end{figure}

The overall trigger scheme is shown in Figure \ref{fg:trigger}. 
The first level trigger requires an incoming kaon 
followed by its decay daughter particle hitting a fiducial
counter and passing through a toroidal magnet gap to the TOF2
counter.
The coincidence logic is defined as
\begin{eqnarray}
{\check C_K} \otimes \sum_{i=1}^{12} ({\rm FID}_i \otimes {\rm TOF2}_i 
\otimes {\rm PL}_i),
\end{eqnarray}
where
$\check C_K$ is the kaon \v Cerenkov counter with multiplicity larger than 6 PMTs hit,
${\rm FID}_i$ is a hit of the {\it i}-th fiducial counter or its
immediate neighbors, and
${\rm TOF2}_i$ and ${\rm PL}_i$ are hits in the TOF2 and polarimeter trigger counters, respectively,  in the {\it i}-th gap.

The first level trigger rate was about 200/beam-spill per gap,
i.e. $2.4\times 10^3$/pulse in total, for $3 \times 10^5$ incident kaons/beam-spill with
a $\pi / K$ ratio of $\sim6$.
The second level trigger required hits in the CsI(Tl) photon detector.
Analog signals from the CsI(Tl) crystals were discriminated
by constant fraction discriminators (CFD) with an energy threshold
of 5 MeV, and then the existence of at least one discriminated signal
within a coincidence time window of 150 ns was required.
Hits in the crystals surrounding the muon hole corresponding 
to the first level trigger vetoed the event in order to suppress  muons 
scattering off the crystals around the muon hole.
A rejection factor at the second level trigger was about 1.7.

The third level trigger required hits in the triple coincidence of 
the positron counter arrays adjacent to
 gap~{\it i} within 20 $\mu$s after the first 
level trigger. 
The rejection factor of the 3rd level trigger is about 13.
Events which did not satisfy either the second or third level triggers by
20 $\mu$s after the first level trigger were cleared.
As a consequence, about 110 events/pulse survived, and 100 events were recorded 
to tape for analysis offline. The break down of the trigger
rates is summarized in Table \ref{tb:trigger}.

      \begin{center}
      \begin{table}
      \caption{Break down of the total 12 gap trigger rate}
      \begin{tabular}{lr}
      \hline
      & $3 \times 10^{12}$ protons/spill \\
      \hline
	1st level trigger & \\
      $\check C_K$ & 300 k/spill \\
	$\check C_K \otimes {\rm FID}$ & 125 k/spill \\
	$\check C_K \otimes {\rm FID} \otimes {\rm TOF}$ & 7.7 k/spill \\
	$\check C_K \otimes {\rm FID} \otimes {\rm TOF}
	 \otimes {\rm PL}$ & 2.4 k/spill \\
	\hline
	1st level $\otimes$ photon & 1.4 k/spill \\
	1st level $\otimes$ photon $\otimes$ positron & 110 /spill\\
	\hline
	Live time fraction = 90\% & 100 /spill\\
      \hline \hline
      \end{tabular}
      \label{tb:trigger}
      \end{table}
      \end{center}

\section{Data Acquisition}
\label{sec:daq}

\subsection{Hardware}
\label{subsec:daqhw}

	The E246 DAQ system was designed as a parallel processing system 
based on  a VME bus, a TKO bus, and standard Fastbus and CAMAC
systems linked with workstations.
The VME bus was used as the main back-end system which collected
all the data using a single-board computer on the VME bus.

 The TKO system was developed and has been used as a standard at KEK\cite{tko}.
It was used to read data from ADC and TDC modules 
connected to the scintillation counter
systems and the MWPCs.

\begin{figure}
\epsfxsize=\linewidth
\begin{center}
\epsffile{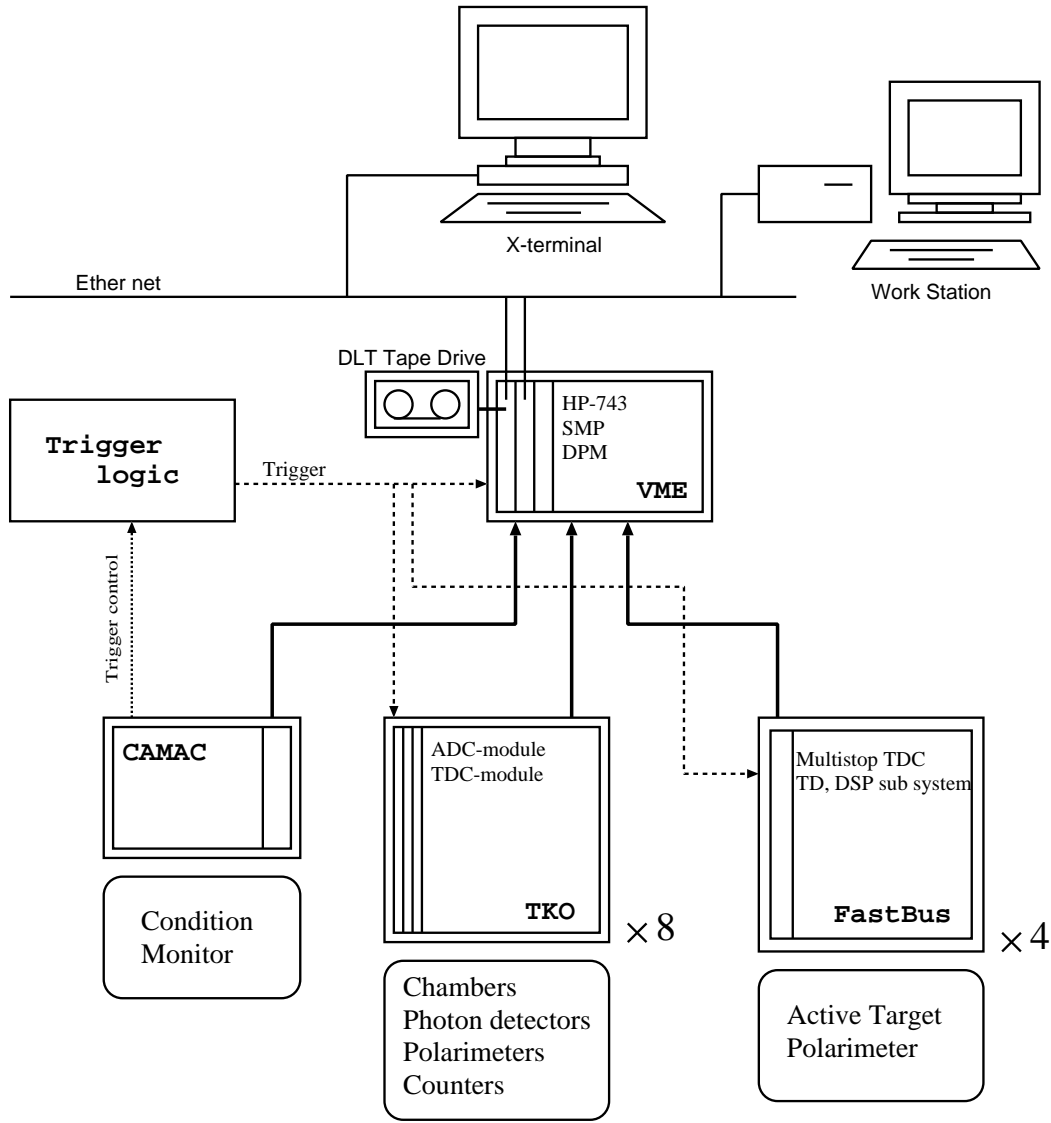}
\end{center}
\caption{A schematic view of the data acquisition (DAQ) configuration.}
\label{fg:daqconfig}
\end{figure}

Fastbus was used to read data from multi-hit TDC modules 
fed from the polarimeter  positron counters and  from the high rate
beam counters.
The SCA transient digitizers for the CsI(Tl)
photon counters were also mounted and read out in Fastbus crates.
CAMAC was used only for slow monitoring and for trigger control.
Workstations were used for online monitors.
The configuration of the DAQ system is shown in Figure \ref{fg:daqconfig}.

TKO and Fastbus have intelligent controllers and buffer memories
and work independently  crate by crate.
	The KEK-PS beam has a spill structure that is 
0.7-s beam on and 2.3-s beam off and
the DAQ system
 took  advantage of this structure.
During the  spill beam-on period, the system only read data from 
the front-end modules and stored the data into local memories in each crate.
During the between spill beam-off period, the stored data were transfered to the 
back-end system, and gathered by the managing process.
The gathered data were stored in a large main buffer which has 
typically about 32 Mbytes capacity and then 
 recorded onto  DLT tape. A part of the data
were distributed to analysis processes via ethernet for 
online monitoring.

\subsubsection{TKO system}
\label{subsubsec:tko}

The TKO intelligent sequencer, the SCH (VME version of the Control Header, CH \cite{tko2}),
has many functions for reading modules and works as a bridge
between TKO and the VME bus.
On the VME bus side, a partner module
called the SMP (S Memory Partner\cite{tko2}), functions 
as a local buffer memory for the TKO data and for controlling the SCH.
A total of eight TKO crates are used to read out the ADCs and TDCs for 
the plastic scintillator PMTs and the CsI(Tl) PIN diodes, as well as the ADCs
for the MWPCs.

\subsubsection{Fastbus system}
\label{subsubsec:fb}

Fastbus  was used to read multi-hit LeCroy 1877/1879 TDC modules
with a time range long enough to cover several  
 2-$\mu$s lifetimes of the muon decay.
The LeCroy 1877/1879 has local buffer memory in each module
which was used in
 pipeline mode.
The conversion time of the multi-hit TDC could
be ignored in this experiment, which was done at a trigger rate of
a few hundred Hz.

To communicate between Fastbus and the VME bus,
we used  a master processor unit in each of the four 
Fastbus crates
which controlled  the 1877 and 1879 modules. This unit has 
a dual-port memory  (DPM) which is connected to both the inner bus and the VME bus
to handle  communication between the two.

A command interpreter for controlling the Fastbus system running
on each unit  received a command from the 
VME system via the DPM to acquire data  independently.
During beam-off this command interpreter
initiated the data transfer from each of the local memories
to the VME bus system also via the DPM.

\subsubsection{CAMAC system}
\label{subsubsec:camac}

 The CAMAC system was used as a trigger control and condition monitor.
In the CAMAC system, output registers, scalers, input registers,
scanning ADCs and a crate controller were used.
The Kinetic Systems 3922 crate controller received commands 
from a partner Kinetics 2940
which resided in the VME crate.
After each beam-on spill, the CAMAC system scanned the 
spill information, trigger conditions,
and counter voltages  for monitoring.

\subsubsection{VME bus system}
\label{subsubsec:vme}

The VME bus system functioned  as master of these systems,
 controlling the other systems and reading the data
collected by each system during the spill.
An HP-753 VME master module (PA-RISC, 64 Mbytes memory) was used 
as the main DAQ controller, running
HP-RT,  a real time operating system based on LynxOS.
This module was connected to a DLT tape drive for data logging.

The slave modules  for the other bus systems that communicated with the
VME bus masters
were eight SMP modules for the TKO system and four AMSK DPM modules for the Fastbus system.
A second HP-753 module on the VME bus running the UNIX-based HP-UX 
 operating system provided the development environment
for the HP-RT system. When the DAQ system was running, this module was available
 to run  online analysis.

\subsubsection{Data recording and online analysis}
\label{subsubsec:onlanalhw}

Two major features of the DLT tape format 
are high recording speed (1.5 Mbytes-s$^{-1}$) and large  capacity (20 Gbytes).
Our data rate was typically 2.4 Mbytes in each  0.7-s long spill every 3.0 seconds for an
 average 0.8 Mbytes-s$^{-1}$, well within the capability of the DLT.

The online analysis mainly for monitoring data quality used
a distributed system over ethernet.
A fraction of the data were transfered via  TCP/IP  to
typically three workstations --- Sun ipx, Sun ipc and intel-PC/Linux.

\subsection{Software}
\label{subsec:softw}

\begin{figure}
\epsfxsize=\linewidth
\begin{center}
\epsffile{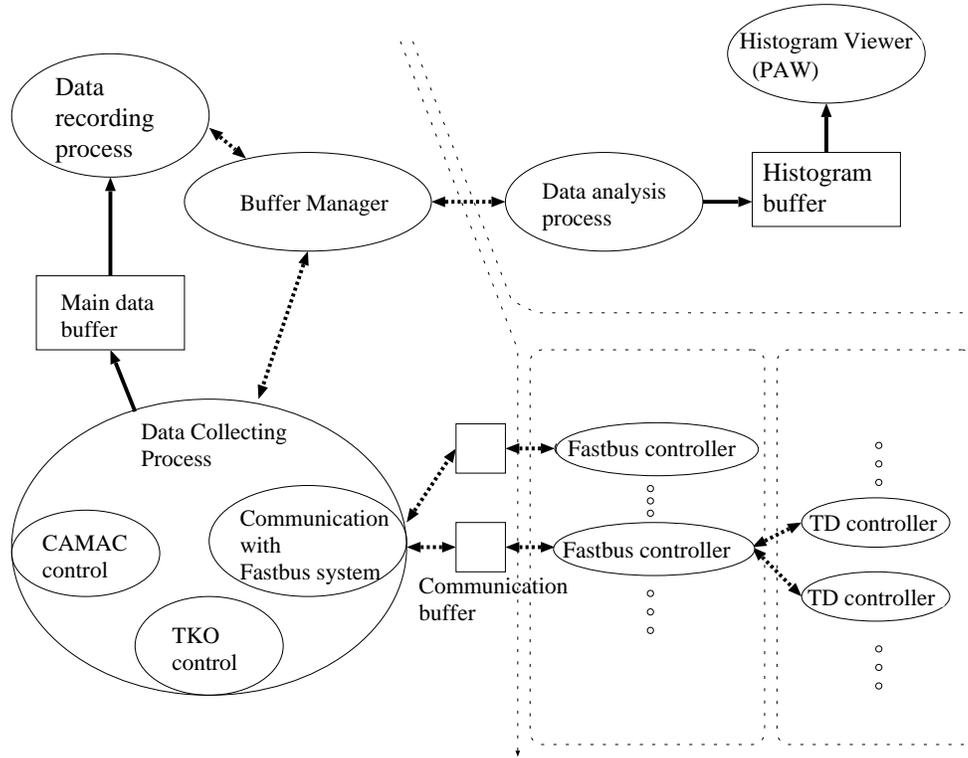}
\end{center}
\caption {Schematic view of the DAQ software.}
\label{fg:daqsoft}
\end{figure}

The DAQ software was based on UNIDAQ\cite{unidaq}, a UNIX-like OS which was developed at KEK 
and  the Tokyo Institute for Technology (TIT).
It includes a buffer manager using
eight 4-Mbyte standard memory data buffers, and a process control system.
The data acquired during one spill were stored
into one buffer.
This DAQ program is an ensemble of single-function modular processes 
which work together:
\begin{itemize}
\item
{\it novad} is a buffer manager which accepts requests from 
the other processes and issues  buffer pointers.
\item
{\it xpc} manages a process database, and checks process status, {\it e.g.}, 
whether the processes are stopped illegally or not.
\item
{\it collector} is a data collector, which controls many devices and
reads data, assembles and packs data, and writes data into a buffer.
\item
{\it operator} is a system control interface for users through which runs are
started and stopped.
\item
{\it recorder} is a data recorder, which writes data from a buffer  onto tape.
\item
{\it analyzer} is an online analysis process, which gets a part of the data
and makes low-level and high-level histograms.
\end{itemize}
A schematic view of the DAQ software is shown in Figure \ref{fg:daqsoft}.

\subsubsection{Collector}
\label{subsubsec:collector}

	The data collection system consists of two programs,
the collector  working on the VME HP-753, and  the CIP
 working on  the Fastbus 68020FPI in each crate.
The collector  communicates with the buffer manager,
controls and collects data from TKO,
 and communicates with the CIP and gets data 
 from Fastbus.
The CIP communicates with the collector, and handles
control and data collection in each Fastbus crate.

\subsubsection{Online analysis}
\label{subsubsec:onlanalsw}

	The online monitor system  is also divided between two programs.
The analyzer gets data from the buffer manager via ethernet
and forms the histograms, and
the other is a histogram viewer.
The analyzer was compiled to run on both Sun UNIX and Linux based computers on the network, requiring Sun compatible IPC
(Internal Process Control) and socket libraries.
The analyzer and histogram viewer worked independently with shared memory.

Viewers  based on PAW/Hbook,
 an analysis tool developed by CERN\footnote{CERN Program Library,
IT division/CERN, http://cernlib.web.cern.ch/cernlib/} and 
on Histo-Scope which was developed by FNAL\footnote{Fermilab Computing Div. Library, http://www-pat.fnal.gov/nirvana/histo.html}, were used.
PAW includes many powerful functions useful for online
analysis, while
Histo-Scope is convenient to use interactively.

\subsection{DAQ system performance}
\label{subsec:daqperf}

	The detector has  about 2000  output channels in total.
One event consisted of about 1600 32-bit words, and at
a typical trigger rate of about 100 to 200 events per spill,
the data size of one spill was about 0.7 to 1.4 Mbytes.

\begin{figure}
\epsfxsize=\linewidth
\begin{center}
\epsffile{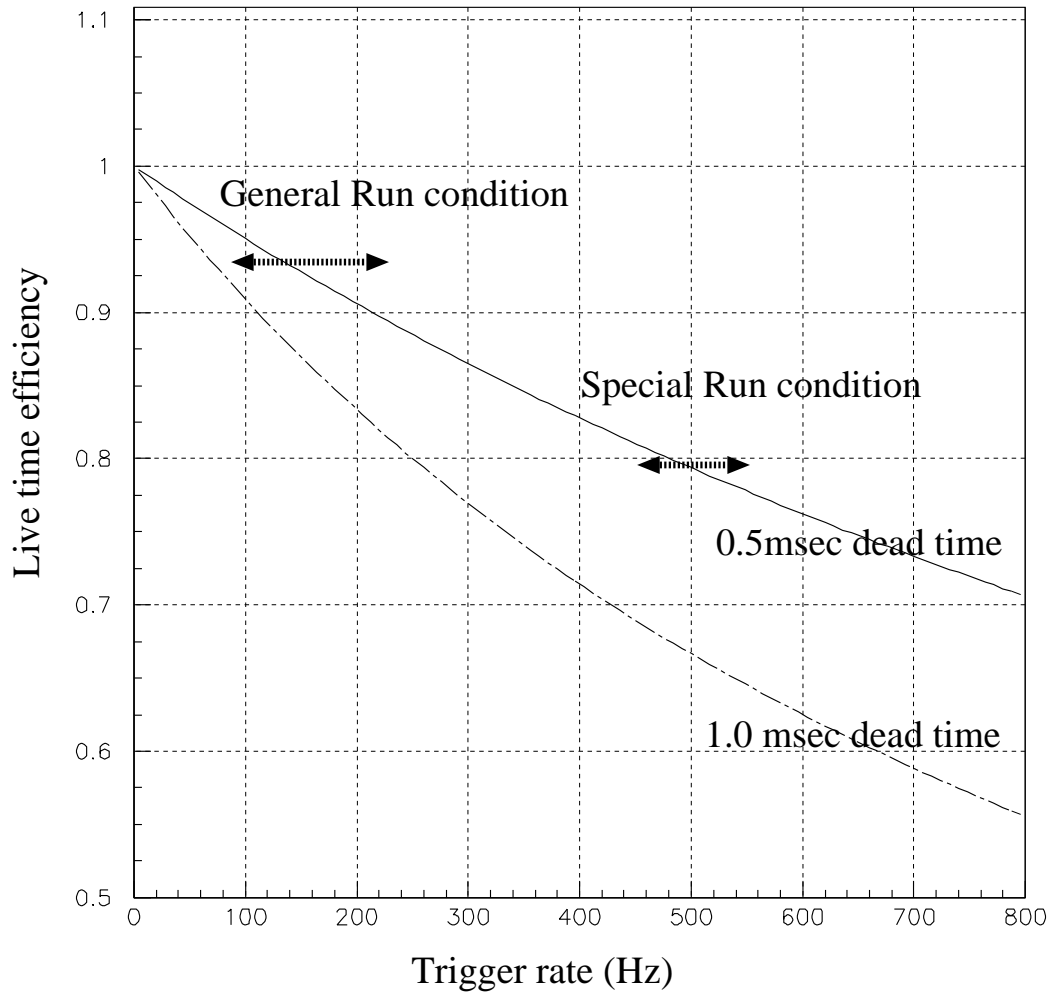}
\end{center}
\caption{DAQ efficiency as a function of trigger rate.}
\label{fg:daqeff}
\end{figure}

The data taking speed  differed in each crate of the TKO system
depending on module configuration.
The trigger timing and the wait for A/D conversion were tuned
individually. The slowest crate needed
150 $\mu$s for conversion and 350 $\mu$s for reading for
 a dead time of 500 $\mu$s.
	The Fastbus system used a pipeline mode
so that A/D conversion time was negligible, but
the readout time was about 500 $\mu$s, so that
the dead time of this system was also about 500 $\mu$s.
The efficiency of the DAQ system depends on the trigger rate and 
the correlation between the trigger rate and the live time efficiency
is shown in Figure \ref{fg:daqeff}.
The DAQ system typically operated with a live time efficiency
from 85\% to 95\%.

\section{Summary}
\label{sec:summary}

The detector built for the search for a T-violating transverse muon polarization in $K_{\mu3}$ decay
at rest included systems for identifying and stopping the low-momentum kaon beam, and precise measurement of the kinematics of  charged decay daughters in a toroidal spectrometer and of photons in a segmented Cs(Tl) calorimeter. The muon polarimeter
used pure aluminum stopping material to avoid depolarization, and used the shimmed fringe 
magnetic field of the superconducting toroid to provide a longitudinal holding field for the azimuthal polarization component. Emphasis on precision component alignment and a high degree of detector symmetry allowed many sources of potential systematic errors to be eliminated or minimized.

For the primary motivating $K_{\mu3}$ T-violation search, systematic errors in the physics analysis have been reduced below the $10^{-3}$ level \cite{fstrslt}. Because of the versatility of the detector, it has also been used to study other $K^+$ decay modes. \cite{papers}
The detector as  described here was dismantled in early 2002.

\section{Acknowledgments}
\label{sec:ack}

This work was supported in Japan by a Grant-in-Aid from the Ministry of 
Education, Science, Sports and Culture, and by JSPS; in Russia by the Ministry 
of Science and Technology, and by the Russian Foundation for Basic Research; in 
Canada by NSERC and IPP, and by infrastructure of TRIUMF provided under its NRC 
contribution; in Korea by BSRI-MOE and KOSEF; in the U.S.A. by  NSF and DOE.

The authors thank S.~Iwata, M.~Kobayashi, V.M.~Lobashev, 
V.A.~Matveev,  K.~Nakai,
K.~Nakamura,
J.-M.~Poutissou, V.A.~Rubakov, H.~Sugawara, E.~Vogt, S.~Yamada and 
T.~Yamazaki for encouragement in executing the present work. They  
gratefully acknowledge the excellent support received from several 
staff members of KEK, INR and TRIUMF, 
in particular, T.~Fujino, E.~Shabalin and N.~Khan. 
They also recognize the early contributions of P.~Bergbusch, 
C.~Chen, D.~Dementyev, M.~Grigoriev,   W.~Keil and I.S.~Park.

\end{document}